\documentclass[pre,aps,floats,twocolumn,floatfix,superscriptaddress]{revtex4}
\usepackage{graphicx,amssymb,amsmath,ifthen,rotating,chemfig}
\usepackage[active]{srcltx}
\begin{document}

\title{Interacting hard-sphere fluids in an external field}

\author{Benaoumeur Bakhti}
\affiliation{G2E Lab, SNV and Department of Physics, University of Mustapha Stambouli, Mascara 29000, Algeria}
\author{Gerhard M\"uller} 
\affiliation{Department of Physics, University of Rhode Island,
  Kingston RI 02881, USA}


\begin{abstract}
We present a new method for studying equilibrium properties of interacting fluids in an arbitrary external field. 
The fluid is composed of monodisperse spherical particles with hard-core repulsion and additional interactions of arbitrary shape and limited range.
Our method of analysis is exact in one dimension and provides demonstrably good approximations in higher dimensions. 
It can cope with homogeneous and inhomogeneous environments.
We derive an equation for the pair distribution function.
The solution,  to be evaluated numerically, in general, or analytically for special cases, enters expressions for the entropy and free energy functionals.
For some one-dimensional systems, our approach yields analytic solutions, reproducing available exact results from different approaches.
\end{abstract}

\maketitle

%
\section{Introduction}\label{sec:sec1} 
%
Fluids of interacting particles, such as sticky or ionic fluids, exhibit broadly varied and complex phenomena, even more so in the inhomogeneous environment of an external field.
Although several experimental works have well characterized the complex behavior of inhomogeneous interacting fluids, a complete microscopic description of the observed behavior is still missing. 
Driven by their widespread applications, including microfluidic devices \cite{Seng13} and nanobiotechnological means of drug delivery \cite{Fanu10}, a great deal of attention has recently been given to models of interacting fluids \cite{MF04a, MF04b, BRP07, HW11}. 

Aiming to better understand equilibrium and nonequilibrium properties down to the nanoscale, a variety of analytical and computational tools have been suggested.
Density functional theory (DFT) is a powerful method for investigating equilibrium properties of interacting fluids \cite{SZK53, Tara85, CA85, Evan92, Loew94, CM97, TK00, TCM08, HM09, Luts10, selgra}. 
However, good approximations from equilibrium DFT have been limited to hard particles (via the fundamental measures approach) or else to weakly interacting particles \cite{Rose89, RSLT96, RSLT97, Tara00, LC02, LC02a, LC04, Roth10}. 
Constructing a good approximation of density functionals for more general interactions remains an open question. 

To a large extent, the complex behavior of real fluids stems from (i) the interaction potentials and (ii) the shapes of constituent particles.
In this work we aim to extend the DFT for fluids of hard bodies beyond the hard-sphere models in the line of aspect (i) by introducing nearest-neighbour interactions of arbitrary profile and coupling strength and with a maximum range of two molecular diameters.

Inspired by work of Percus on hard rods \cite{Perc89} and by the probabilistic modeling of lattice hard rods with additional interactions, we derive a recurrence relation for the pair distribution function (PDF) of interacting hard-sphere fluids which is valid in any dimension \cite{RV81, BMD00, BMD00a, BSM12, BMM13, BKMM15}.
Using this recurrence relation, we infer explicit expressions for entropy and free energy as functionals of density and PDF.
A key advantage of our approach is that the total correlation function or the
radial distribution function can be determined directly from the PDF without
further differentiation of the free energy functional, which circumnavigates the most laborious parts of more commonly taken approaches.
The close connection between radial distribution functions and neutron scattering cross sections makes the former a desirable object of theoretical and computational investigations.

\begin{figure}[t!]
\begin{center}
\includegraphics[width=35mm,angle=-90]{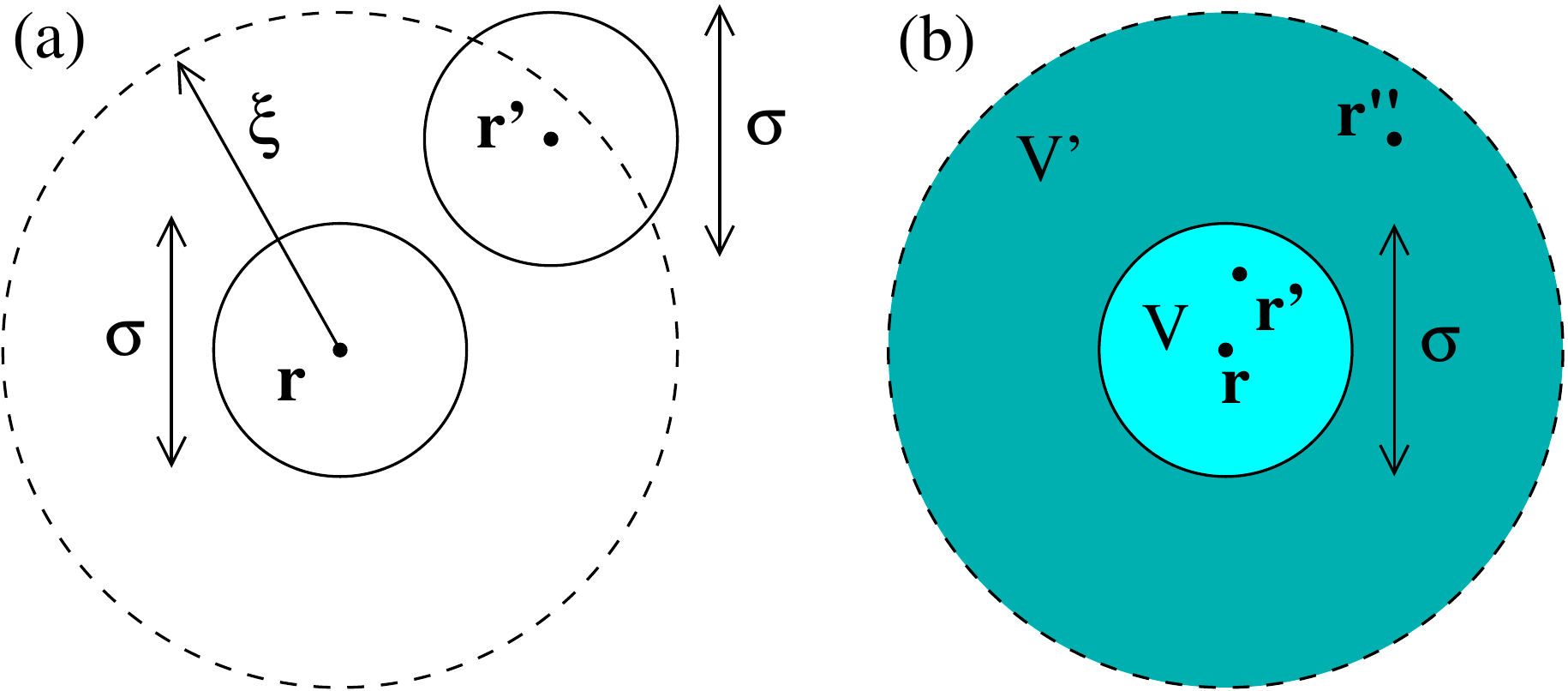}
\end{center}
\caption{ (a) The pair interaction, $\epsilon(|\mathbf{r}-\mathbf{r}'|)$, between two hardcore particles of diameter $\sigma$ has range, $\sigma\leq|\mathbf{r}-\mathbf{r'}|\leq\xi$.
(b) Regions of integration, named $V$ and $V'$, are shaded differently. Position $\mathbf{r}$ is at the center of $V$, position $\mathbf{r}'$ anywhere inside $V$ and position $\mathbf{r}''$ anywhere inside $V'$.}
\label{fig:colloid}
\end{figure}

In this work we consider a fluid consisting of $N$ identical particles confined to some macroscopic region of space, in the presence of an external potential $V_{ex}$ and a pair interaction potential $\phi$ between neighbors within a limited range of mutual distances. 
The total (potential) energy is \cite{note1, note2},
\begin{align}\label{eq:4b}
\mathcal{H}(\mathbf{r}_1,\ldots,\mathbf{r}_N) = \sum_{i<j}\phi(\mathbf{r}_i,\mathbf{r}_j) + \sum_{i=1}^{N}V_{ex}(\mathbf{r}_i).
\end{align} 
The particles have a hard core of diameter $\sigma$.
They are rods, disks, and spheres in $\mathcal{D}=1, 2, 3$ dimensions, respectively.
In addition to hardcore repulsion, our model includes a central-force pair interaction $\epsilon(r)$ of limited range and arbitrary profile.
We thus write,
\begin{equation}\label{eq:1}
\phi(\mathbf{r}_i,\mathbf{r}_j)= \left\{
\begin{array}{ll}
\infty & \quad :~ |\mathbf{r}_i-\mathbf{r}_j| < \sigma, \\
\epsilon\big(|\mathbf{r}_i-\mathbf{r}_j|\big)   & \quad :~ 
\sigma\leq |\mathbf{r}_i-\mathbf{r}_j| < \xi, \\
0 & \quad :~ |\mathbf{r}_i-\mathbf{r}_j| \geq \xi.
\end{array} \right.
\end{equation}
The restriction, $\xi<2\sigma$, for the maximum interaction range in combination with the hardcore repulsion [Fig.~\ref{fig:colloid}(a)] ensures that the number of neighbors to which any particle is coupled is limited by the number of nearest-neighbors in a close-packed configuration: two in $\mathcal{D}=1$, six in $\mathcal{D}=2$ (hexagonal), and twelve in $\mathcal{D}=3$ (hcp or fcc).

In Sec.~\ref{sec:sec2} we present the central idea behind the methodology -- a recurrence relations for the PDF -- that produces exact results in $\mathcal{D}=1$ and approximations of promising quality in higher $\mathcal{D}$. 
The practical use of this recurrence relation in the functionals for the entropy and the free energy are described in Sec.~\ref{sec:entropy_free_energy} for very general situations.
In applications to 1D systems, all expressions simplify considerably as is described in Sec.~\ref{sec:sec4}, and the approach is exact as is demonstrated in Appendix~\ref{sec:appa}.
The strength of the methodology for 3D applications is shown in Sec.~\ref{sec:sec5} in the context of three situations with inhomogeneous external potentials.

%
\section{Pair distribution function}\label{sec:sec2} 
%
Direct correlation functions (DCF) are of central interest in DFT.
Generally, the DCF can be determined from the free energy functional via second derivative with respect to the particle density. 
If the free energy functional is not known, it can be determined from the Ornstein-Zernike (OZ) equation. 
But here a closure relation is required between the two unknowns of the OZ equation, namely the DCF and the total correlation function. 
Different closure approximations have been proposed to solve the OZ equation, including the Percus-Yevick \cite{PY58}, hypernetted chain \cite{LGB59, RA79}, Born-Green \cite{BG46} and mean-spherical-model approximations \cite{LP66}. 
With the DCFs thus calculated, free-energy functionals in various approximations are then being constructed and evaluated.

The route we take in this work is different and does not rely on any closure relation.
It aims instead for a hitherto unused relation between the particle density,
\begin{align}\label{eq:2c}
\rho(\mathbf{r})\doteq\left\langle\hat{\rho}(\mathbf{r})\right\rangle
=\frac{1}{N}\left\langle\sum_{i=1}^N\delta(\mathbf{r}-\mathbf{r}_i)\right\rangle.
\end{align}
and the PDF \cite{note3},
\begin{align}\label{eq:2a}
\rho^{(2)}(\mathbf{r},\mathbf{r'})
&=\left\langle\hat{\rho}(\mathbf{r})\hat{\rho}(\mathbf{r'})\right\rangle 
\nonumber \\
&=\frac{1}{N(N-1)}\left\langle\sum_{i,j\neq i}\delta(\mathbf{r}-\mathbf{r}_i)\delta(\mathbf{r'}-\mathbf{r}_j) \right\rangle,
\end{align}
where $\langle\hspace{0.9mm}\rangle$ denotes the canonical ensemble average.

For the establishment of this relation between $\rho(\mathbf{r})$ and $\rho^{(2)}(\mathbf{r},\mathbf{r'})$, we introduce three auxiliary distribution functions (ADFs) defined as follows:
\begin{align}\label{eq:2f}
\tilde{\rho}(\mathbf{r},\mathbf{r'})\doteq
\left\langle\hat{\rho}(\mathbf{r'})\left[1-\int_{V'}d\bar{\mathbf{r}}_2\hat{\rho}(\bar{\mathbf{r}}_2)\right] \right\rangle,
\end{align}
\begin{align}\label{eq:2g}
\tilde{\rho}_1(\mathbf{r},\mathbf{r''})\doteq
\left\langle\left[1-\int_{V}d\bar{\mathbf{r}}_1\hat{\rho}(\bar{\mathbf{r}}_1)\right]\hat{\rho}(\mathbf{r''})\right\rangle,
\end{align}
\begin{align}\label{eq:2h}
\tilde{\rho}_0(\mathbf{r})\doteq\left\langle\!\left[1-\!\!\int_{V}d\bar{\mathbf{r}}_1\hat{\rho}(\bar{\mathbf{r}}_1)\right]\!\!\left[1-\!\!\int_{V'}d\bar{\mathbf{r}}_2\hat{\rho}(\bar{\mathbf{r}}_2)\right]\! \right\rangle,
\end{align}
where the regions of integration are explained in Fig.~\ref{fig:colloid}(b).
The variable $\mathbf{r}$ is only implicitly present in these expressions.
The positions $\mathbf{r}'$ and $\mathbf{r}''$ relative to the position $\mathbf{r}$ are limited to the regions $V$ and $V'$.

These expressions can be interpreted as the probability density (\ref{eq:2f}) for a particle to be present at position $\mathbf{r'}$ in region $V$ and no particle present inside region $V'$, the probability density (\ref{eq:2g}) for a particle to be present at position $\mathbf{r''}$ in region $V'$ and no particle present inside region $V$, and the probability density (\ref{eq:2h}) for no particle to be present inside the combined region $V_T\doteq V\cup V'$.

The ADFs are not independent of each other.
All three can be expressed as functionals of the density (\ref{eq:2c}) and the PDF (\ref{eq:2a}):
\begin{align}
&\tilde{\rho}(\mathbf{r},\mathbf{r'})=\rho(\mathbf{r'})-\int_{V'}d\bar{\mathbf{r}}_2 \rho^{(2)}(\mathbf{r'},\bar{\mathbf{r}}_2),\label{eq:rho_tilde_1}\\
&\tilde{\rho}_1(\mathbf{r},\mathbf{r''})=\rho(\mathbf{r''}) - \int_{V}d\bar{\mathbf{r}}_1\rho^{(2)}(\bar{\mathbf{r}}_1,\mathbf{r''}),\label{eq:rho_tilde_2}\\
&\tilde{\rho}_0(\mathbf{r})=1 - \int_{V_T}d\bar{\mathbf{r}}\rho(\bar{\mathbf{r}}) \label{eq:rho_tilde_3}\\
&\hspace{15mm}+ \int_{V}d\bar{\mathbf{r}}_1\int_{V'}d\bar{\mathbf{r}}_2\rho^{(2)}(\bar{\mathbf{r}}_1,\bar{\mathbf{r}}_2).\nonumber
\end{align}
There is space for no more than one particle in region $V$.
If a particle is present in region $V$, then there is space for no more than one particle in region $V'$.

Next we relate the PDF and the three ADFs to the joint probability distribution (JDF) at thermal equilibrium,
\begin{align}\label{eq:4a}
p(\mathbf{r}_1,\ldots,\mathbf{r}_N)=Z^{-1}e^{-\beta\mathcal{H}(\mathbf{r}_1,\ldots,\mathbf{r}_N)},
\end{align}
where
\begin{equation}\label{eq:4aa}
Z=\int_\mathcal{V} d\mathbf{r}_1\cdots \int_\mathcal{V}d\mathbf{r}_N
e^{-\beta\mathcal{H}(\mathbf{r}_1,\ldots,\mathbf{r}_N)},
\end{equation}
is the canonical partition function and $\mathcal{V}$ the space to which the fluid is confined.
Keeping in mind that here we are using a canonical ensemble, the four distribution functions can be represented as follows:
\begin{subequations}\label{eq:5a-1}
\begin{align}
\rho^{(2)}(\mathbf{r'},\mathbf{r''})= &\frac{1}{Z}\frac{1}{N(N-1)}
\nonumber \\
&\hspace*{-18mm}\times e^{-\beta[\phi(\mathbf{r'},\mathbf{r''})+V_{ex}(\mathbf{r'})+V_{ex}(\mathbf{r''})]}
\int_\mathcal{V'}\!\! d^{N-2}\mathbf{r}\,e^{-\beta\mathcal{H}_1},
\end{align}
\begin{align}
&\mathcal{H}_1(\mathbf{r}',\mathbf{r''},\mathbf{r}_3,\ldots,\mathbf{r}_N)= 
\nonumber \\
&\hspace{20mm}\mathcal{H}-\phi(\mathbf{r'},\mathbf{r''})-V_{ex}(\mathbf{r'})-V_{ex}(\mathbf{r''}),
\end{align}
\end{subequations}
\begin{align}\label{eq:5a-4}
\tilde{\rho}_0(\mathbf{r})= \frac{1}{Z}\frac{1}{N(N-1)}\!\!\int_\mathcal{V'}\!\!d^{N}\mathbf{r}\,e^{-\beta\mathcal{H}},
\end{align}
\begin{subequations}\label{eq:5a-2}
\begin{align}
\tilde{\rho}(\mathbf{r},\mathbf{r'})=\frac{1}{Z}\frac{1}{N(N-1)}e^{-\beta V_{ex}(\mathbf{r'})}\!\!\int_\mathcal{V'}\!\! d^{N-1}\mathbf{r}e^{-\beta\mathcal{H}_2},
\end{align}
\begin{align}
\mathcal{H}_2(\mathbf{r}',\mathbf{r}_2,\ldots,\mathbf{r}_N)= 
\mathcal{H}-V_{ex}(\mathbf{r'}),
\end{align}
\end{subequations}
\begin{subequations}\label{eq:5a-3}
\begin{align}
\tilde{\rho}_1(\mathbf{r},\mathbf{r''})= \frac{1}{Z}\frac{1}{N(N-1)}e^{-\beta V_{ex}(\mathbf{r''})}\!\!\int_\mathcal{V'} \!\!d^{N-1}\mathbf{r}e^{-\beta\mathcal{H}_3}
\end{align}
\begin{align}
\mathcal{H}_3(\mathbf{r}_1,\mathbf{r''},\mathbf{r}_3,\ldots,\mathbf{r}_N)= 
\mathcal{H}-V_{ex}(\mathbf{r''}), 
\end{align}
\end{subequations}
where $\mathcal{V'}$ is the complement to $V_T$ in $\mathcal{V}$: $\mathcal{V}=\mathcal{V'}\cup V_T$, $\mathcal{V'}\cap V_T=\emptyset$.
We can thus write,
\begin{align}\label{eq:5}
\frac{\tilde{\rho}_0(\mathbf{r})\rho^{(2)}(\mathbf{r'},\mathbf{r''})}
{\tilde{\rho}(\mathbf{r},\mathbf{r'})\tilde{\rho}_1(\mathbf{r},\mathbf{r''})}
=e^{-\beta\phi(\mathbf{r'},\mathbf{r''})}
\frac{A(\mathbf{r},\mathbf{r'},\mathbf{r''})}{B(\mathbf{r},\mathbf{r'},\mathbf{r''})},
\end{align}
where\\ \\
\begin{equation}\label{eq:5A}
A(\mathbf{r},\mathbf{r'},\mathbf{r''})=
\left[\int_\mathcal{V'}\!\!d^{N}\mathbf{r}\,e^{-\beta\mathcal{H}}\right]
\left[\int_\mathcal{V'}\!\! d^{N-2}\mathbf{r}\,e^{-\beta\mathcal{H}_1}\right],
\end{equation}
\begin{equation}\label{eq:5B}
B(\mathbf{r},\mathbf{r'},\mathbf{r''})=
\left[\int_\mathcal{V'}\!\! d^{N-1}\mathbf{r}e^{-\beta\mathcal{H}_2}\right]
\left[\int_\mathcal{V'} \!\!d^{N-1}\mathbf{r}e^{-\beta\mathcal{H}_3}\right].
\end{equation}

In what follows we shall use Eq.~(\ref{eq:5}) for $A=B$ with the understanding that, in general, it represents an approximation. 
We shall argue that the approximation is good, in general.
Indeed, there are nontrivial situations for which $A=B$ is exact as shown in Appendix~\ref{sec:appa}.
In combination with expressions (\ref{eq:5a-1})-(\ref{eq:5a-3}) , the assumption $A=B$  yields the following functional relation -- the desired recurrence relation -- between the particle density and the PDF:
\begin{widetext}
\begin{align}\label{eq:6}
\rho^{(2)}(\mathbf{r'},\mathbf{r''})=e^{-\beta\phi(\mathbf{r'},\mathbf{r''})}
\frac{\Bigl[\rho(\mathbf{r'}) - \int_{V'} d\bar{\mathbf{r}}_2\rho^{(2)}(\mathbf{r'},\bar{\mathbf{r}}_2)\Bigr]\Bigl[\rho(\mathbf{r''}) - \int_V d\bar{\mathbf{r}}_1\rho^{(2)}(\bar{\mathbf{r}}_1,\mathbf{r''})\Bigr]}{\Bigl[1 - \int_{V_T} d\bar{\mathbf{r}}\rho(\bar{\mathbf{r}}) + \int_{V} d\bar{\mathbf{r}}_1\int_{V'}d\bar{\mathbf{r}}_2\rho^{(2)}(\bar{\mathbf{r}}_1,\bar{\mathbf{r}}_2)\Bigr]}\quad :~ 
\sigma\leq |\mathbf{r}_i-\mathbf{r}_j| < \xi.
\end{align}
\end{widetext}
For distances outside this range we have,
\begin{equation}\label{eq:6plus}
\rho^{(2)}(\mathbf{r'},\mathbf{r''})=\left\{
\begin{array}{ll} 0 & :~ |\mathbf{r}_i-\mathbf{r}_j|<\sigma, \\
\rho(\mathbf{r'})\rho(\mathbf{r''}) & :~ |\mathbf{r}_i-\mathbf{r}_j|>\xi.
\end{array} \right.
\end{equation}
For specific 1D models this implicit relation can be solved into an explicit expression for the PDF as a functional of the particle density.
In general, we must solve Eq.~(\ref{eq:6}) computationally.

The numerical results shown below in Sec.~\ref{sec:sec5} justify the strength of the approximation $A=B$. For interacting systems,
the highly structured oscillations appearing in density profiles can be captured only by an advanced DFT approximation such as the fundamental measure theory (FMT), but not by a simple DFT approximation such as the local density approximation (a mean field theory) and can only be partially captured by the weighted density approximation.

The radial distribution function and total correlation function follow directly:
\begin{equation}
g(\mathbf{r'},\mathbf{r''})=\frac{\rho^{(2)}(\mathbf{r'},\mathbf{r''})}{\rho(\mathbf{r'})\rho(\mathbf{r''})},\quad h(\mathbf{r'},\mathbf{r''})=g(\mathbf{r'},\mathbf{r''})-1.
\end{equation}
Even though the derivation of (\ref{eq:6}) has been worked out in the canonical ensemble, the result is independent of the ensemble in use.
It will be used in the grandcanonical ensemble, to which we switch in Sec.~\ref{sec:sec3}. 

%
\section{Free-energy and entropy functionals}\label{sec:entropy_free_energy}\label{sec:sec3} 
%
DFT expresses the free energy (grand potential) as a functional, 
\begin{equation}\label{eq:22} 
\tilde{\Omega}=\tilde{\Omega}[\tilde{V}_{ex},\phi],
\end{equation}
of the external potential modified by the chemical potential,
\begin{equation}\label{eq:23} 
\tilde{V}_{ex}(\mathbf{r})\doteq \mu-V_{ex}(\mathbf{r}),
\end{equation}
and the interaction potential $\phi(\mathbf{r},\mathbf{r'})$ such as introduced in (\ref{eq:1}).
The density $\rho(\mathbf{r})$ and the PDF $\rho^{(2)}(\mathbf{r},\mathbf{r'})$ are extracted from (\ref{eq:22}) via functional derivatives:
\begin{align}
\rho(\mathbf{r}) =-\frac{\delta\tilde{\Omega}[\tilde{V}_{ex},\phi]}{\delta\tilde{V}_{ex}},\quad 
\rho^{(2)}(\mathbf{r},\mathbf{r'}) &=\frac{\delta\tilde{\Omega}[\tilde{V}_{ex},\phi]}{\delta\phi}. \label{eq:pdf_int_pot}
\end{align}

The functions $\rho(\mathbf{r})$ and $\rho^{(2)}(\mathbf{r},\mathbf{r'})$ are conjugate to the functions $\tilde{V}_{ex}(\mathbf{r})$ and $\phi(\mathbf{r},\mathbf{r'})$, respectively, in a thermodynamic sense.
Performing a Legendre transform on (\ref{eq:22}) yields the expression \cite{Perc89},
\begin{widetext}
\begin{align}\label{eq:legendre-transf}
&\int d\mathbf{r}\,\tilde{V}_{ex}(\mathbf{r})\,\frac{\delta\tilde{\Omega}}{\delta\tilde{V}_{ex}(\mathbf{r})} 
+ \int d\mathbf{r} \int d\mathbf{r'}\,\phi(\mathbf{r},\mathbf{r'})\,\frac{\delta\tilde{\Omega}}{\delta\phi(\mathbf{r},\mathbf{r'})} 
- \Omega\nonumber\\
&= -\mu\int d\mathbf{r}\rho(\mathbf{r})
+\int d\mathbf{r}\,V_{ex}(\mathbf{r})\rho(\mathbf{r})
+\int d\mathbf{r} \int d\mathbf{r'}\,\phi(\mathbf{r},\mathbf{r'})
\rho^{(2)}(\mathbf{r},\mathbf{r'}) -\tilde{\Omega}
= TS[\rho,\rho^{(2)}].
\end{align}
The last equation in (\ref{eq:legendre-transf}), which produces the entropy functional, is evident if we identify the first term as $-G$ (Gibbs free energy) and the sum of the next two terms as $U$ (internal energy).
For what follows, we introduce two kinds of entropy density functionals by writing,
\begin{equation}\label{eq:entr-dens-func} 
S[\rho,\rho^{(2)}]=\int d\mathbf{r}\int d\mathbf{r'}\,
\tilde{S}[\rho(\mathbf{r}),\rho^{(2)}(\mathbf{r},\mathbf{r'})],\quad
\bar{S}[\rho,\rho^{(2)}]=\int d\mathbf{r'}\,
\tilde{S}[\rho(\mathbf{r}),\rho^{(2)}(\mathbf{r},\mathbf{r'})].
\end{equation}

The free-energy as a functional of density and PDF now reads,
\begin{align}\label{eq:7}
\tilde{\Omega}=\Omega[\rho(\mathbf{r}),\rho^{(2)}(\mathbf{r},\mathbf{r'})]
 = \int d\mathbf{r} d\mathbf{r'} \rho^{(2)}(\mathbf{r},\mathbf{r'})\phi(\mathbf{r},\mathbf{r'}) + \int d\mathbf{r}\,\rho(\mathbf{r})V_{ex}(\mathbf{r})
- TS[\rho,\rho^{(2)}]
- \mu\int d\mathbf{r}\,\rho(\mathbf{r}).
\end{align}
\end{widetext}
The external potential and the interaction potential can be extracted from the entropy functional as,
\begin{align}
\tilde{V}_{ex}(\mathbf{r}) =-T\frac{\delta \bar{S}[\rho,\rho^{(2)}]}{\delta\rho}, \quad
\phi(\mathbf{r},\mathbf{r'}) =T\frac{\delta\tilde{S}[\rho,\rho^{(2)}]}{\delta\rho^{(2)}}. \label{eq:int_pot_entropy}
\end{align}

An alternative route to Eqs.~(\ref{eq:int_pot_entropy}) invokes extremum conditions for the free-energy functional (\ref{eq:7}):
\begin{align}\label{eq:fe_min}
\frac{\delta\Omega[\rho,\rho^{(2)}]}{\delta\rho}=0,\quad
\frac{\delta\Omega[\rho,\rho^{(2)}]}{\delta\rho^{(2)}}=0,
\end{align}
previously employed in different contexts by Gonis et al. \cite{GSET96, GSTE97} and Bakhti \cite{Bakh13}.
Carrying out the operations in Eqs.~(\ref{eq:fe_min}) using expressions (\ref{eq:7}) indeed reproduces the results of  Eqs.~(\ref{eq:int_pot_entropy}).

Equations~(\ref{eq:int_pot_entropy}) state that if the entropy functional is known, the external potential and the interaction potential which generate certain profiles for density and PDF can be calculated uniquely. 
Conversely, if the external potential and the interaction potential are known, the density and the PDF can be determined by integrating Eqs.~(\ref{eq:int_pot_entropy}).
We can thus start from the second Eq.~(\ref{eq:int_pot_entropy}) and continue our analysis by inferring from it the relation (see Appendix~\ref{sec:appb}),
\begin{equation}\label{eq:8}
T\tilde{S}[\rho,\rho^{(2)}] = \int\!\!\phi(\mathbf{r},\mathbf{r'})\hspace{0.5mm}d\rho^{(2)}(\mathbf{r},\mathbf{r'}).
\end{equation}

In order to perform this integral, we rewrite the recurrence relation (\ref{eq:6}) in the form,
\begin{widetext}
\begin{align}\label{eq:32}
\beta\phi(\mathbf{r},\mathbf{r'})
=-\ln\Bigl[1 - \int d\mathbf{r}_1\rho(\mathbf{r}_1) &+ \int d\mathbf{r}_1d\mathbf{r'}_1\rho^{(2)}(\mathbf{r}_1,\mathbf{r'}_1)\Bigr]
-\ln\rho^{(2)}(\mathbf{r},\mathbf{r'}) \nonumber \\
&+\ln\Bigl[\rho(\mathbf{r}) - \int d\mathbf{r'}\rho^{(2)}(\mathbf{r},\mathbf{r'})\Bigr]
+\ln\Bigl[\rho(\mathbf{r'}) - \int d\mathbf{r}\rho^{(2)}(\mathbf{r},\mathbf{r'})\Bigr]
\end{align}
The integral (\ref{eq:8}) can now be calculated without further approximations.
It produces an explicit expression for the entropy as a functional of density and PDF:
\begin{align}\label{eq:9}
S[\rho,\rho^{(2)}]/k_B 
&=\int d\mathbf{r}\Bigl\{-\Bigl[\rho(\mathbf{r}) -\int d\mathbf{r'}\rho^{(2)}(\mathbf{r},\mathbf{r'})\Bigr]\ln\Bigl[\rho(\mathbf{r}) -\int d\mathbf{r'}\rho^{(2)}(\mathbf{r},\mathbf{r'})\Bigr] \nonumber\\
&-\Bigl[\rho(\mathbf{r}) - \int d\mathbf{r'}\rho^{(2)}(\mathbf{r'},\mathbf{r})\Bigr]\ln\Bigl[\rho(\mathbf{r}) - \int d\mathbf{r'}\rho^{(2)}(\mathbf{r'},\mathbf{r})\Bigr] 
 - \Bigl[1 - \int d\mathbf{r}_1\rho(\mathbf{r}_1) + \int d\mathbf{r}_1d\mathbf{r'}_1\rho^{(2)}(\mathbf{r}_1,\mathbf{r}_1')\Bigr] \nonumber \\
&\times\ln\Bigl[1 - \int d\mathbf{r}_1\rho(\mathbf{r}_1)+ \int d\mathbf{r}_1d\mathbf{r}_1'\rho^{(2)}(\mathbf{r}_1,\mathbf{r}_1')\Bigr] - \int d\mathbf{r'}\rho^{(2)}(\mathbf{r},\mathbf{r'})\ln\rho^{(2)}(\mathbf{r},\mathbf{r'})\Bigr\}+ \int d\mathbf{r}\rho(\mathbf{r})\ln\rho(\mathbf{r}),
\end{align}
where the last term is an ``integration constant'', added to accommodate the noninteracting limit, $\phi(\mathbf{r},\mathbf{r'})\to0$.
When we now combine Eqs.~(\ref{eq:7}), (\ref{eq:32}),  and (\ref{eq:9}), the free energy functional acquires the form,
\begin{align}\label{eq:12}
\beta\Omega[\rho,\rho^{(2)}] &=\int d\mathbf{r}\Bigl\{\rho(\mathbf{r})\ln\Bigl(\rho(\mathbf{r}) - \int d\mathbf{r'}\rho^{(2)}(\mathbf{r},\mathbf{r'})\Bigr)
+\rho(\mathbf{r})\ln\Bigl(\rho(\mathbf{r}) -\int\! d\mathbf{r'}\rho^{(2)}(\mathbf{r'},\mathbf{r})\Bigr) 
\nonumber \\
&+\Bigl[1 - \int_V d\mathbf{r}_1\rho(\mathbf{r}_1)\Bigr] 
\ln\Bigl(1- \int_V d\mathbf{r}_1\rho(\mathbf{r}_1) + \int d\mathbf{r}_1d\mathbf{r'}_1\rho^{(2)}(\mathbf{r}_1,\mathbf{r}_1')\Bigr) 
- \rho(\mathbf{r})\ln\rho(\mathbf{r}) - \beta\tilde{V}_{ex}(\mathbf{r})\rho(\mathbf{r})\Bigr\}.
\end{align}
\end{widetext}
In the limit, $\phi(\mathbf{r},\mathbf{r'})\to0$, expression (\ref{eq:12}) neatly reduces to the free-energy functional of the hard-sphere model. 
The full expression is consistent with the FMT functionals,
reflecting the contributions from the one particle cavity (terms with $\rho$ only), and contributions from the two-particle cavity (terms with $\rho$ and $\rho^{(2)}$) \cite{LC05}.

Any thermodynamic function and response function of interest can be inferred from (\ref{eq:12}). 
The expressions developed with this methodology are valid for 3D systems.
In general, the path to explicit results requires that we resort to a numerical analysis.
Three applications to colloidal systems in a heterogeneous, 3D  environment are presented in Sec.~\ref{sec:sec5}. 
Further applications are in the works \cite{Bakhti-et-al}.

For 3D sticky-core fluids, the interaction range is reduced to the hard-sphere diameter $(\xi\rightarrow \sigma)$. 
In consequence, the volume integrals $V'$ in Eq~ (\ref{eq:6}) or (\ref{eq:32}) have to be replaced by  surface integrals $S$ over spheres of diameter $\sigma$. 
The regions $S$ and $V$ represent the surface and the interior space of the sphere with diameter $\sigma$ centered at position $\mathbf{r}$, respectively. 
The vectors $\mathbf{r}$ and $\mathbf{r'}$ are related by $|\mathbf{r}-\mathbf{r'}|=\sigma$.
Expressions (\ref{eq:9}) and (\ref{eq:12}) for the entropy and free-energy functionals must be adapted accordingly.

We conclude Sec.~\ref{sec:sec3} by reiterating that our approach bypasses the OZ equation. 
The extremum of the free-energy functional produces one relation between the particle density $\rho(\mathbf{r})$ and the PDF $\rho^{(2)}(\mathbf{r},\mathbf{r'})$. A second relation is Eq.~(\ref{eq:6}), which, in general, represents an approximation. 
The latter is inferred from (\ref{eq:5}), which is exact, but does not lead to closure, except under special circumstances such as discussed in Sec.~\ref{sec:sec4} and Appendix~\ref{sec:appa}. 
The OZ formalism deals with the same problem differently.
In both methods, achieving closure comes, in general, at the cost of approximation. 

%
\section{Exact analysis for 1D systems}\label{sec:sec4} 
%
The entropy and free-energy expressions developed above are amenable to an exact analysis for 1D systems with specific interactions.
Consider a system of hard rods of length $\sigma$ confined to a channel and with   
the interaction potential (\ref{eq:1}) left unspecified for now.
Equation~(\ref{eq:6}) then reads
\begin{widetext}
\begin{align}\label{eq:13}
\rho^{(2)}(y,y')=e^{-\beta\phi(y,y')}\,
\frac{\Bigl[\rho(y) - \int dy'_1\,\rho^{(2)}(y,y'_1)\Bigr]\Bigl[\rho(y') - \int dy'_1\,\rho^{(2)}(y'_1,y')\Bigr]}{\Bigl[1 - \int dy'_1\,\rho(y'_1) + \int dy'_1\int dy'_2\,\rho^{(2)}(y'_1,y'_2)\Bigr]}
\end{align}
The range of $y'$ appearing in the integral of the PDF consists of two intervals: $[y-\xi/2,y-\sigma/2]$ and $[y+\sigma/2,y+\xi/2]$.
The entropy expression (\ref{eq:9}) acquires the form,
\begin{align}\label{eq:14}
\frac{S[\rho,\rho^{(2)}]}{k_B} &=
-\int dy\!\left\lbrace\Bigl[\rho(y) - \int dy'\rho^{(2)}(y,y')\Bigr]\ln\Bigl(\rho(y) - \int dy'\rho^{(2)}(y,y')\Bigr)\right.
\nonumber \\
&+\Bigl[\rho(y) - \int dy'\rho^{(2)}(y',y)\Bigr]\ln\Bigl(\rho(y) - \int dy'\rho^{(2)}(y',y)\Bigr) 
 -\rho(y)\ln\rho(y) +\int dy'\rho^{(2)}(y,y')\ln\rho^{(2)}(y,y')
\nonumber\\
&+\Bigl[1 - \int dy_1\rho(y_1) + \int dy_1dy'_1\rho^{(2)}(y_1,y'_1)\Bigr]
\left.\ln\Bigl(1 - \int dy_1\rho(y_1) + \int dy_1dy'_1\rho^{(2)}(y_1,y'_1)\Bigr)\right\rbrace.
\end{align}
Our approach also produces an exact expression for the free energy, 
\begin{align}\label{eq:15}
\beta\Omega[\rho,\rho^{(2)}] &=\int dy\Bigl\{\rho(y)\ln\Bigl(\rho(y) - \int dy'\rho^{(2)}(y,y')\Bigr)
+\rho(y)\ln\Bigl(\rho(y) - \int dy'\rho^{(2)}(y',y)\Bigr)
\nonumber \\
&+ \Bigl[1 - \int dy_1\rho(y_1)\Bigr]
\ln\Bigl(1 - \int dy_1\rho(y_1) + \int dy_1dy'_1\rho^{(2)}(y_1,y'_1)\Bigr) 
- \rho(y)\ln\rho(y)  -\beta\tilde{V}_{ex}(y)\rho(y)\Bigr\},
\end{align}
\end{widetext}
derived from expression (\ref{eq:12}) as a special case of much wider scope, albeit not exact in higher dimensions.

The entropy expression (\ref{eq:14}) coincides exactly with the result (2.21) of Percus in \cite{Perc89}. 
Given the completely different nature of the two approaches, this is a remarkable convergence.
In addition to the mathematical interest that exact solutions draw quite generally, in the present context they are also of practical interest.
Within the DFT formalism, exact solutions for 1D systems present themselves as ingredients in the construction of approximate functionals in higher dimensions.
Moreover, 3D systems that are inhomogeneous in only one direction are often modeled (with tacit caveats) as effectively 1D systems.

Our approach has a remarkable simplicity for what it is capable to deliver. 
It certainly is much simpler than the inverse-problem approach.
There is a straightforward path to extend our method to situations with general nearest-neighbor and next-nearest-neighbor interactions.
This can be accomplished by introducing three-point auxiliary distribution functions.
Contact can then be made with  the inverse-problem approach of Percus on a wider scope  \cite{BP96, Perc97, Wert84, Wert86}.
One strength of our approach is that the higher-dimensional functionals are constructed without the FMT reference to dimensional crossover and zero-dimensional cavity. 

The functional relation (\ref{eq:13}) between density and PDF can be solved explicitly for $\rho^{(2)}(y,y')$ if we restrict the interaction to a sticky-core contact interaction, where we have $\xi=\sigma$.
We can then express (\ref{eq:13}) in the form,
\begin{widetext}
\begin{align}\label{eq:16}
\rho^{(2)}(y,y')=e^{-\beta\phi(y,y')}\,
\frac{\Bigl[\rho(y) - \rho^{(2)}(y,y')\Bigr]\Bigl[\rho(y') - \rho^{(2)}(y,y')\Bigr]}{1 - L(y) + \rho^{(2)}(y,y')},\quad 
L(y)\doteq \int_{y-\sigma/2}^{y+\sigma/2}dx\,\rho(x),
\end{align}
and rewrite it as a quadratic equation for the PDF,
\begin{align*}
-\left[1-L(y)+e^{-\beta\phi(y,y')}(\rho(y)+\rho(y'))\right]\rho^{(2)}(y,y')
+(e^{-\beta\phi(y,y')}-1)\Bigl[\rho^{(2)}(y,y')\Bigr]^{2}+ e^{-\beta\phi(y,y')}\rho(y)\rho(y')=0
\end{align*}
which has the unique physically relevant solution,
\begin{align}\label{eq:17}
\rho^{(2)}(y,y')=
\frac{1}{2\eta}\Bigl[K[\rho]- \sqrt{K^{2}[\rho] - 4\eta(\eta + 1)\rho(y)\rho(y')}\Bigr],\quad 
K[\rho]\doteq 1+e^{-\beta\phi(y,y')}\Bigl[\rho(y)+\rho(y')\Bigr]- L(y),
\end{align}
where $\eta = e^{-\beta\phi(y,y')}-1$ and $y'$ is now related to $y$ by $y'-y=\sigma$.
The simplified expressions for the entropy (\ref{eq:14})  and for the free energy (\ref{eq:15}), into which we can substitute the PDF (\ref{eq:17}), read
\begin{align}\label{eq:19}
\frac{S[\rho,\rho^{(2)}]}{k_B} &=
\int dy\Bigl\{-\Bigl[\rho(y) - \rho^{(2)}(y,y')\Bigr]\ln\Bigl(\rho(y) - \rho^{(2)}(y,y')\Bigr) 
-\Bigl[\rho(y') - \rho^{(2)}(y,y')\Bigr]\ln\Bigl(\rho(y') - \rho^{(2)}(y,y')\Bigr) \nonumber\\
& \hspace*{-5mm}-\Bigl[1 - L(y) + \rho^{(2)}(y,y')\Bigr]\ln\Bigl(1 - L(y) + \rho^{(2)}(y,y')\Bigr)
 - \rho^{(2)}(y,y')\ln\Bigl(\rho^{(2)}(y,y')\Bigr) +\rho(y)\ln\rho(y)\Bigr\},
\end{align} 
\begin{align}\label{eq:20}
\beta\Omega[\rho,\rho^{(2)}]&=\int dy\left\{\rho(y)\ln\Bigl(\rho(y) - \rho^{(2)}(y,y')\Bigr)
+\rho(y')\ln\Bigl(\rho(y') - \rho^{(2)}(y,y')\Bigr)\right.\nonumber\\
&\left.\hspace{15mm}+\Bigl[1 - L(y)\Bigr]\ln\Bigl(1 - L(y) + \rho^{(2)}(y,y')\Bigr)
- \rho(y)\ln\rho(y) - \beta\tilde{V}_{ex}(y)\rho(y)\right\}.
\end{align}
We are now ready to demonstrate the strength of our approach by showcasing two inhomogeneous model systems. 
\vspace{5mm}
\end{widetext}

%
\section{Applications}\label{sec:sec5} 
%
First we investigate  the sticky-hard-sphere (SHS) model in the presence of an attractive Lennard-Jones (LJ) wall. 
Second we investigate the effect of gravity on the SHS system. 
These two systems are of great importance for studying wetting transitions, sedimentation, and interfacial phenomena in interacting colloids. 
Our focus is on density profiles pertaining to the 3D case.

A more systematic study of this system, including effects of longer-range interactions and lower dimensionality, presently in the works \cite{Bakhti-et-al}, is of great importance for studying glass transition under gravity. 
The SHS model \cite{Baxter:1968} has proven to be realistic for many physical phenomena including crystallization of polymers \cite{Hoy/OHern:2010}, micelles \cite{Amokrane/Regnaut:1997}, protein solutions \cite{Braun:2002}, DNA coated colloids \cite{Xu/etal:2011, Dreyfus/etal:2009}, and ionic fluids \cite{Stell:1995}.  
In the SHS model, the interaction between colloidal particles is limited to an adhesive force upon contact. 

It is well known that a significant attractive interaction between colloids results from the depletion forces, which come into play when polymer globules or micelles are added to a colloidal suspension.  
The range and strength of the attraction can be varied continuously and independently by adjusting, respectively, the concentration and size of the polymer.  
For weak depletion, the colloidal particles are well described by the SHS model.

Consider the square-well interaction potential,
\begin{align}\label{eq:sw_potential}
\beta\phi(r)=\left\lbrace\begin{array}{ll}
+\infty & :~ 0<r<\sigma,\\
-\ln\left[\sigma/12\tau(\sigma-\xi)\right] & :~ \sigma<r<\xi,\\
0 & :~ r>\xi,
\end{array}\right.
\end{align}
of width $\xi$, where $\tau$ has been named Baxter temperature.. 
In the limit SHS limit $\xi\rightarrow\sigma$ we can write,
\begin{align}\label{eq:shs_boltz}
e^{-\beta\phi(r)}=\theta(r-\sigma)+\frac{\sigma}{12\tau}\delta(r-\sigma).
\end{align}
Noninteracting hard spheres are recovered for $\tau\rightarrow \infty$.

\subsection{SHS model with LJ adhesive wall}\label{sec:sec5a} 
Here we consider the effect of an attractive planar wall on an SHS system.
We construct an attractive external potential via
\begin{align}
V_{ext}(z)&=E_w\int_{-\infty}^{\infty}dx'\int_{-\infty}^{\infty}dy'\int_{-\infty}^{0}dz'\nonumber\\
&\hspace{10mm}\times \phi_\mathrm{LJ}\left(\sqrt{x'^2+y'^2+(z-z')^2}\,\right)
\end{align}
from the LJ interaction,
\begin{align}\label{eq:LJ_potential}
\phi_\mathrm{LJ}(r)=\left\lbrace\begin{array}{ll}
0 & :~ r\leq \sigma_w,\\
\displaystyle 4\epsilon\left[\left(\frac{\sigma_w}{r}\right)^{12}-\left(\frac{\sigma_w}{r}\right)^6\right] & :~ r>\sigma_w,
\end{array}\right.
\end{align}
where $\epsilon$ is the interaction strength.
Setting $\sigma_w=\sigma$ and performing the integral yields a an effective external potential of the form,
\begin{align}\label{eq:plan_wall_LJ}
V_{ex}(z)=E_w\left[-\frac{1}{6}\left(\frac{\sigma}{z}\right)^3+\frac{1}{45}\left(\frac{\sigma}{z}\right)^9\right],
\end{align}
where $E_w$ is an energy parameter and $z$ the distance from the wall. 

An external potential of the type (\ref{eq:plan_wall_LJ}) was previously introduced as a simplified model of substrate potential for alkaline metals \cite{Ebner/etal:1976,Yatsyshin/etal:2012}, notwithstanding the fact that there exist more accurate models \cite{Marshall/etal:1996,Chizmeshya/etal:1998}. 
The unidirectional nature of the external potential reduces the 3D SHS problem technically to a 1D problem, the results of planar averaging \cite{Yatsyshin/etal:2012,Jamnik:1998,Rodriguez/Vicente:1996}.
Such quasi-1D structures are, of course, very different from the structures that characterize truly 1D systems.

From Eqs.~(\ref{eq:int_pot_entropy}) we derive the following general relation between external potential, density, and PDF:
\begin{align}\label{eq:ddim_mu}
\beta[V_{ex}(\mathbf{r})-\mu]=
&\ln\Bigl[1 - \!\int\!\! d\mathbf{r}_1\rho(\mathbf{r}_1) + \!\!\int\!\! d\mathbf{r}_1d\mathbf{r'}_1\rho^{(2)}(\mathbf{r}_1,\mathbf{r'}_1)\Bigr]\nonumber\\
&+\ln\rho(\mathbf{r})-\ln\Bigl[\rho(\mathbf{r}) - \!\!\int\!\! d\mathbf{r'}\rho^{(2)}(\mathbf{r},\mathbf{r'})\Bigr]\nonumber\\
&-\ln\Bigl[\rho(\mathbf{r}) - \int\!\! d\mathbf{r'}\rho^{(2)}(\mathbf{r'},\mathbf{r})\Bigr].
\end{align}
This equation in combination with Eq.~(\ref{eq:32}), which brings the interaction potential into play, determine the density and the PDF.
We solved them numerically for the SHS interaction, using the Newton and the Broyden algorithms. 
The chemical potential and the temperature can be tuned independently to fix the average bulk (reservoir) density via the relation,
\begin{align}\label{eq:rho_normalization}
\frac{1}{L}\int_0^L\rho(z)=\rho_b,
\end{align}
where $\rho_b$ is the reservoir density. 
For consistency with notation found in the literature, we define the average density 
\begin{align}\label{eq:rho_average}
\bar{\rho}=\rho\sigma^3
\end{align}
and control it with the parameter $\bar{\mu}=\beta\mu$.

\begin{figure}[t]
\begin{center}
\includegraphics[scale=0.28]{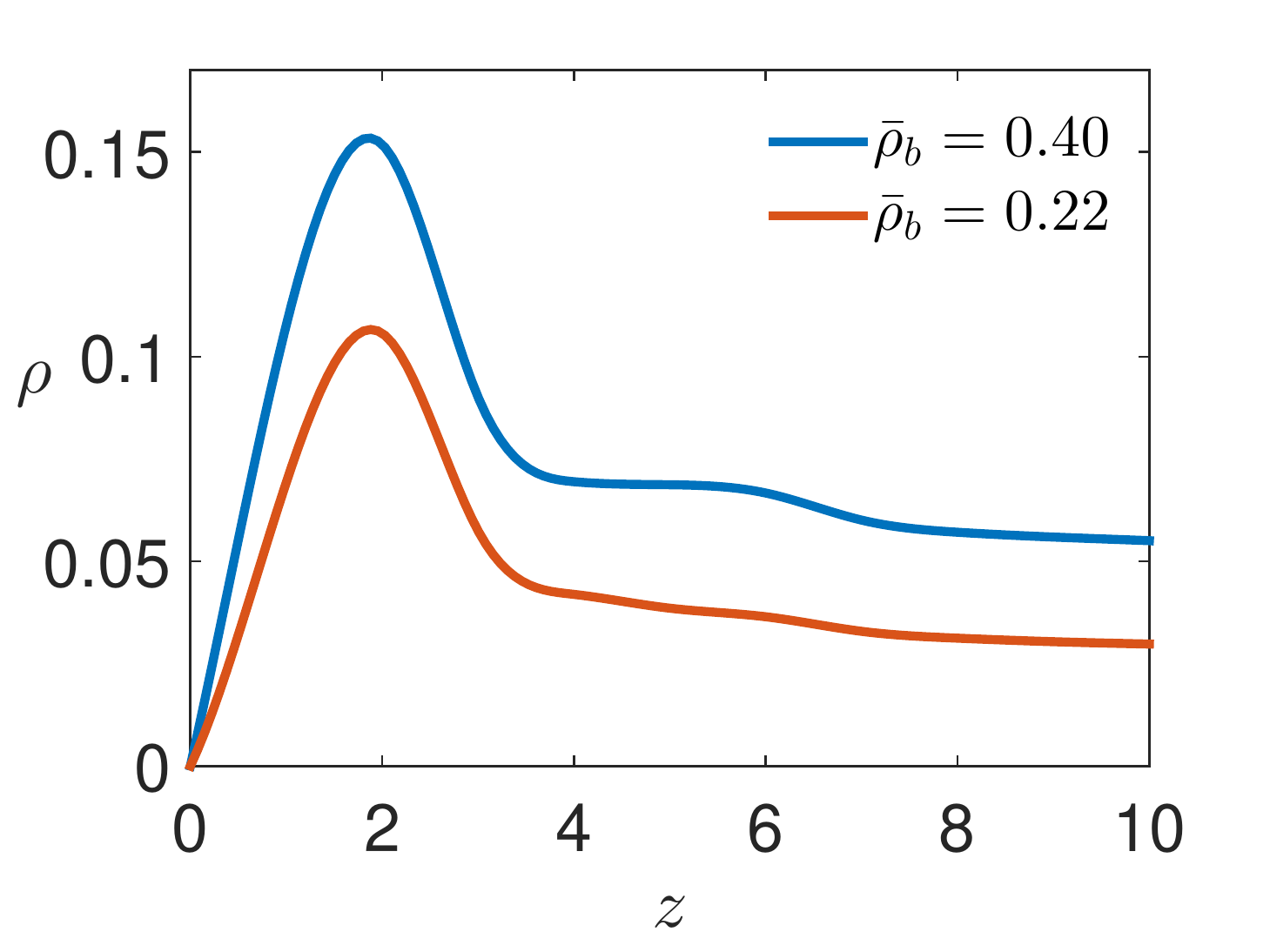}
\includegraphics[scale=0.28]{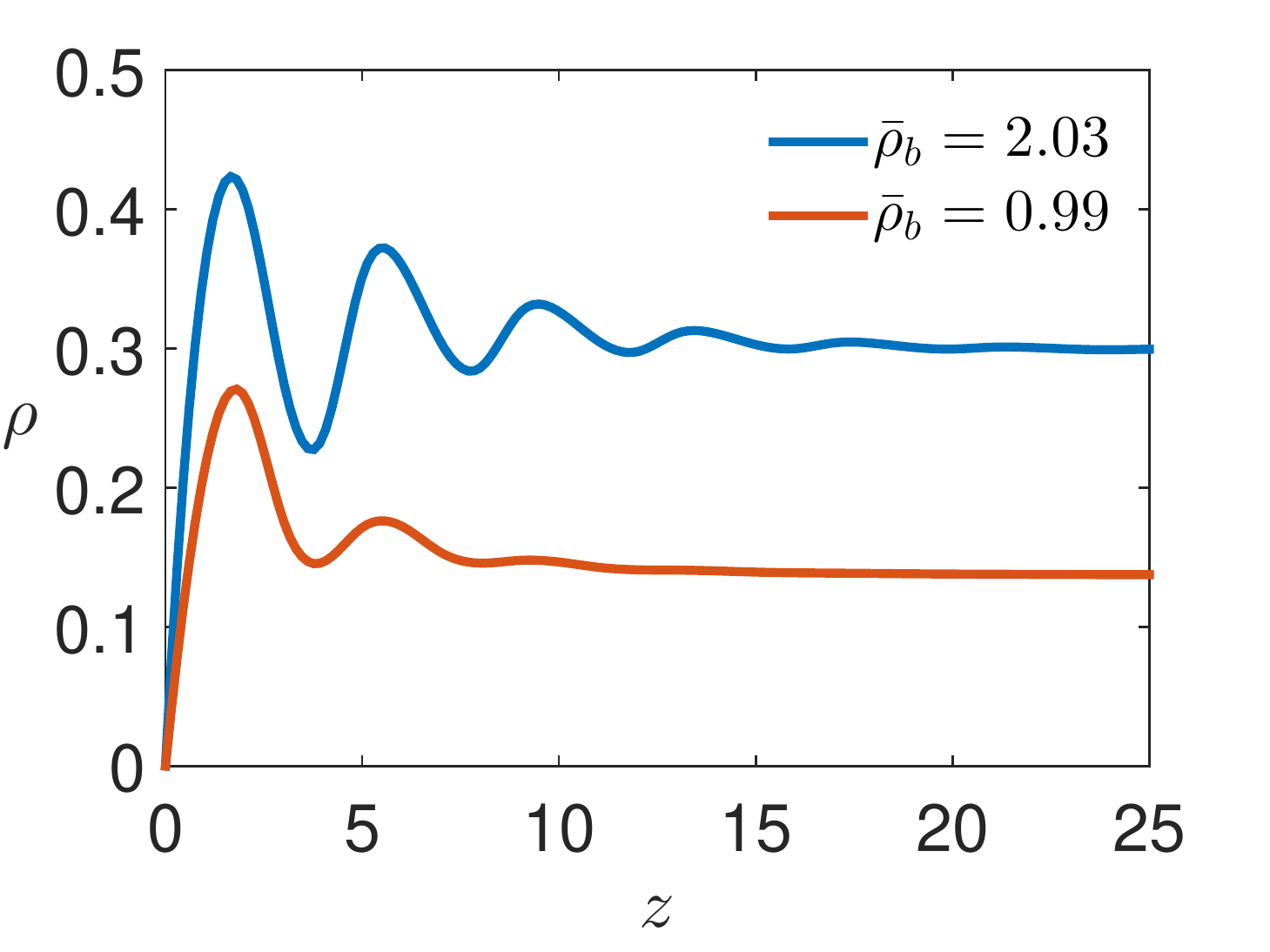}
\end{center}
\caption{Density profiles near LJ wall for parameter values $\sigma=4$, $\tau=0.7$, $E_w=8.0$.
The four curves pertain to average bulk densities $\bar{\rho}_b=0.22$, $0.40$, $0.99$, $2.03$, controlled by the parameter values $\bar{\mu}=-3.6$, $-3.0$, $-2.7$ and $-1.5$, respectively.}
\label{fig:rho_T}
\end{figure}

\begin{figure}[b]
\begin{center}
\includegraphics[scale=0.28]{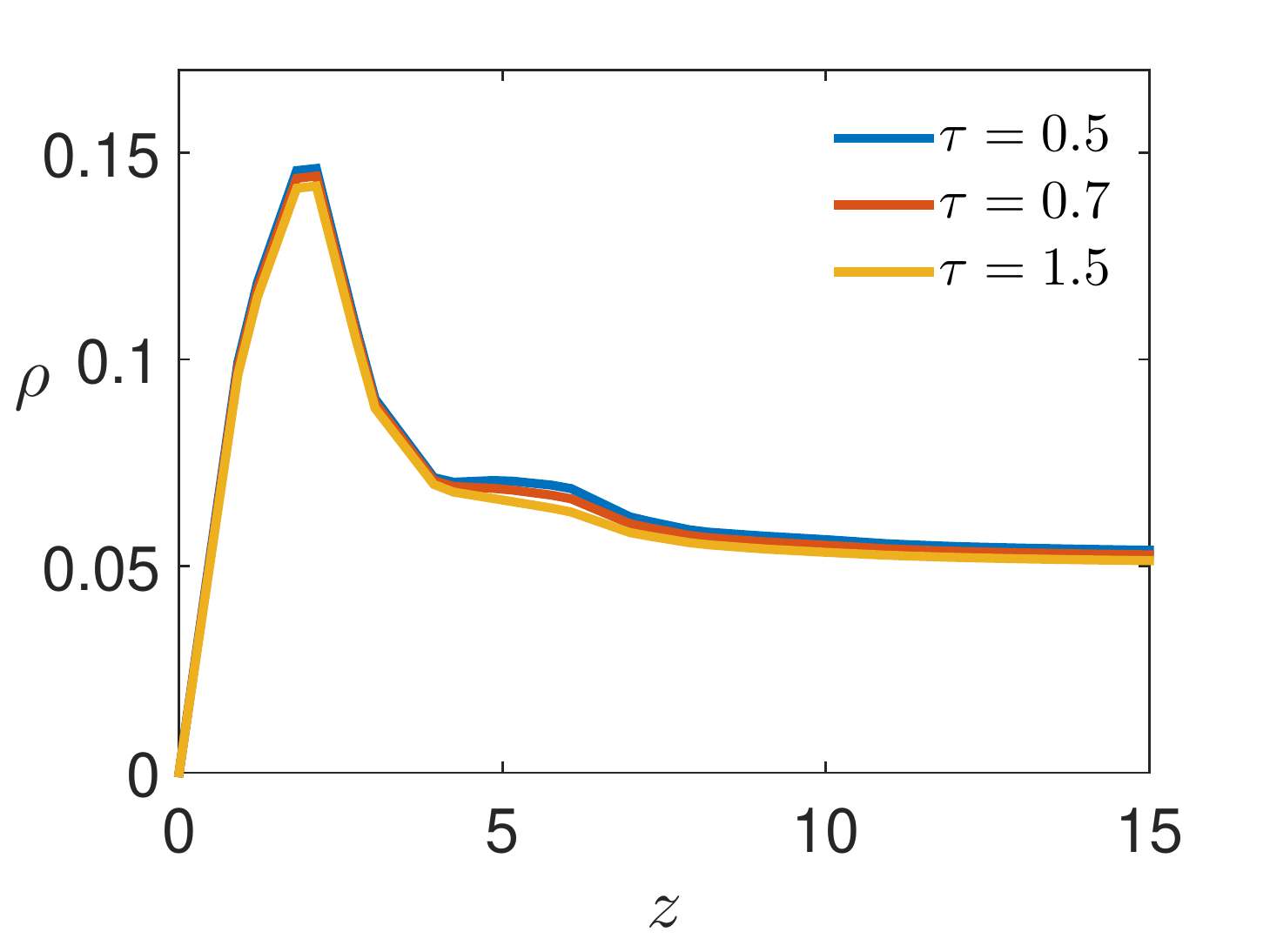}
\includegraphics[scale=0.28]{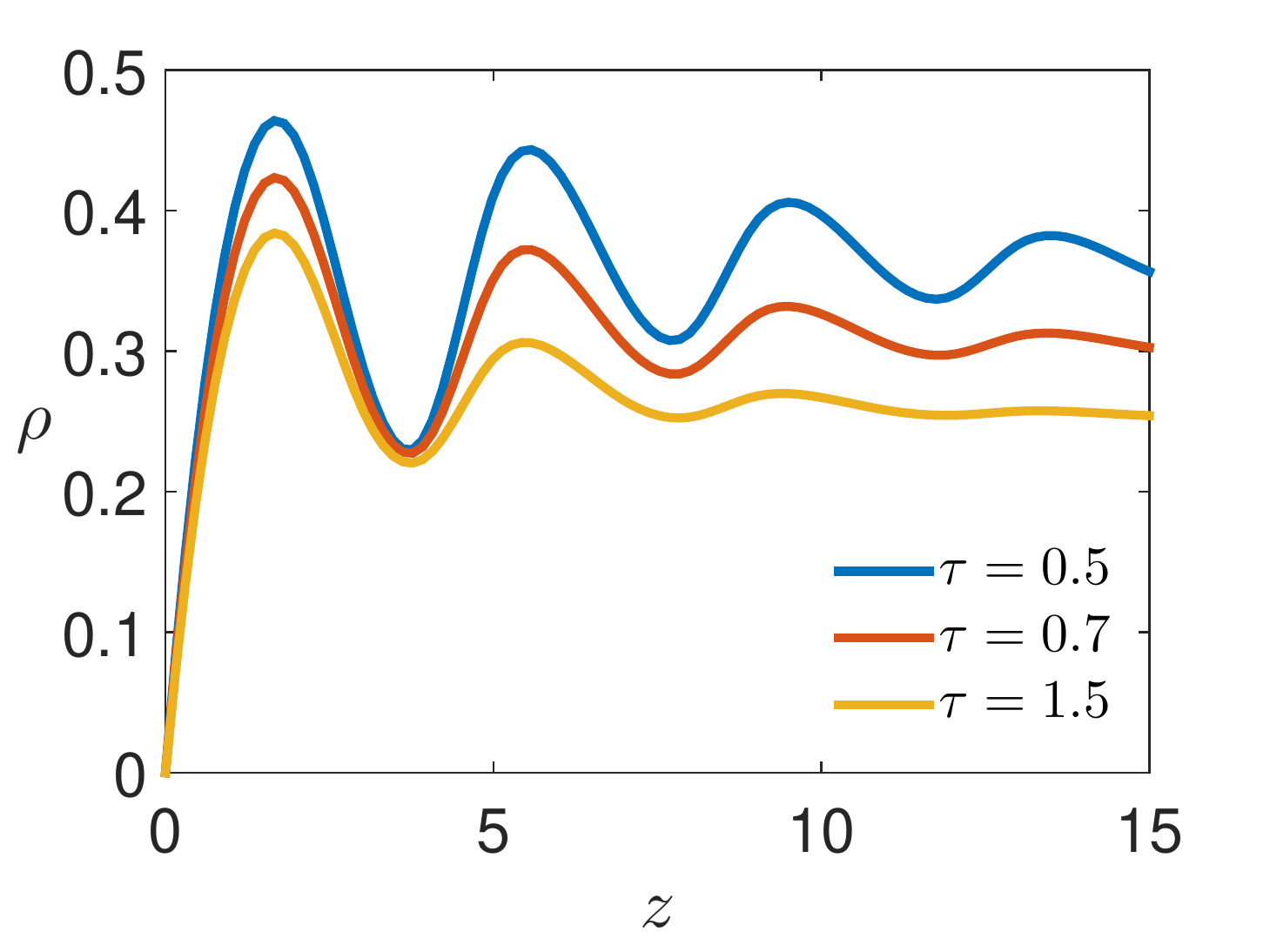}
\end{center}
\caption{Density profiles near LJ wall at $k_BT=1.0$ (a) and $k_BT=0.5$ (b) for parameter values $\sigma=4$, $\mu=-3$ and $E_w=8.0$, and various strengths of stickiness.}
\label{fig:rho_tau}
\end{figure}

\begin{figure}[t]
\begin{center}
\includegraphics[scale=0.50]{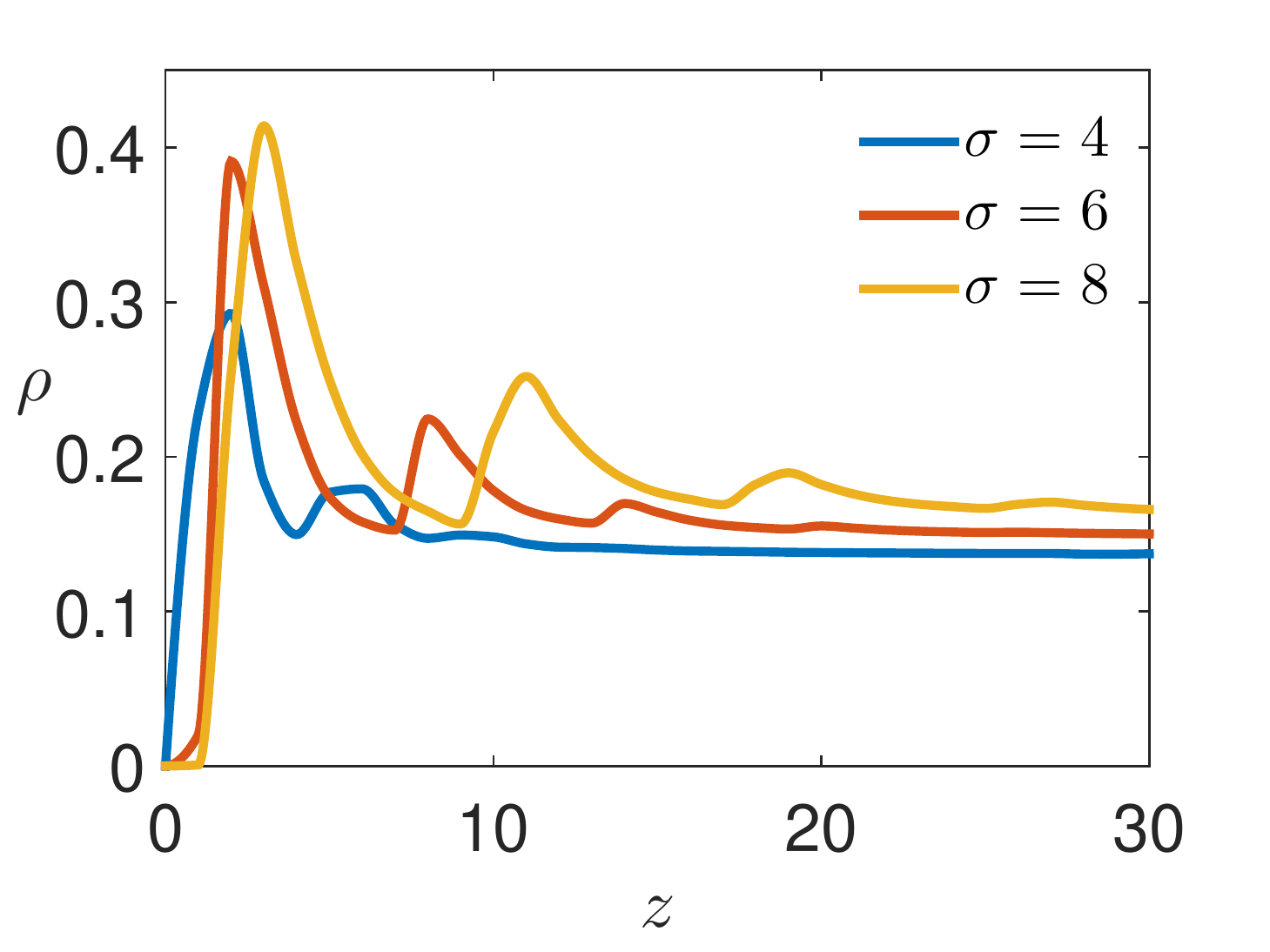}
\end{center}
\caption{Density profiles near LJ wall for parameter values $\tau=0.7$, $\beta = 0.6$, $\mu=-3$ and $E_w=9$, and particles of various diameters.}
\label{fig:rho_sigma}
\end{figure}

The numerical results presented in Figs.~\ref{fig:rho_T}-\ref{fig:rho_sigma} showcase different aspects of the results. 
Figure~\ref{fig:rho_T} shows density profiles at different values of the average reservoir density $\bar{\rho}_b$. 
We observe the emergence of oscillations near the wall as $\bar{\rho}_b$ increases.
They are the combined effect of wall attraction and hard-sphere repulsion.
Such highly structured oscillations cannot be captured by DFT using the local density approximation (LDA) or the weighted density approximation (WDA).
It requires more accurate functionals such as provided by FMT.

The dependence of the density profile on the stickiness parameter at two different temperatures is shown in Fig.~\ref{fig:rho_tau}. 
We see that the stickiness of the particle surface makes a difference only at temperatures sufficiently low that the adhesive energy between two particles prevails against the thermal energy $k_BT$.
Larger particles produce oscillations near the wall that have a larger amplitude, a larger wavelength, and a larger depth into the bulk, as is evident in the results presented in Fig.~\ref{fig:rho_sigma}.

\subsection{SHS model with hard walls}\label{sec:sec5c} 
\begin{figure}[h]
\begin{center}
\includegraphics[scale=0.28] {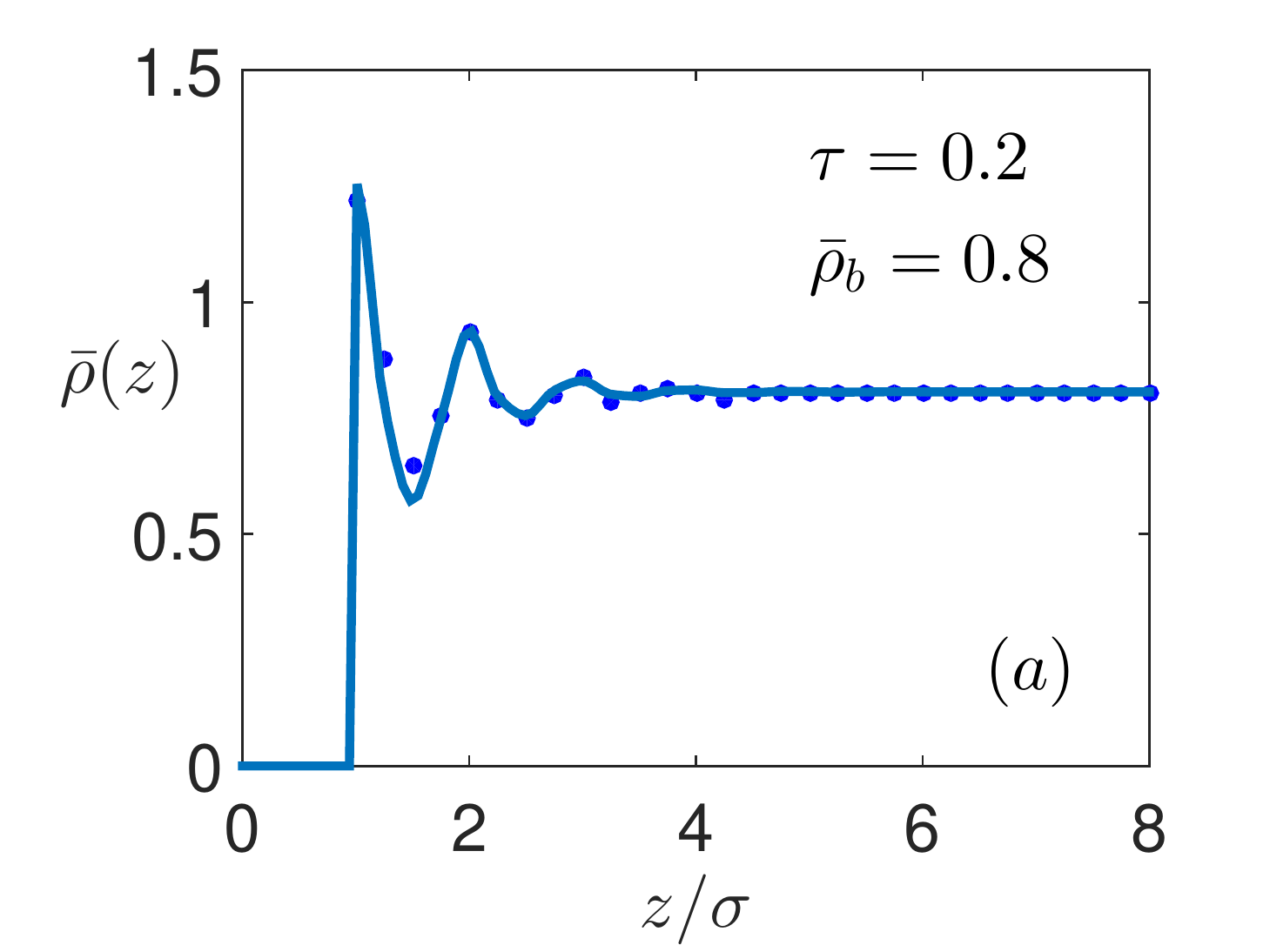}
\includegraphics[scale=0.28] {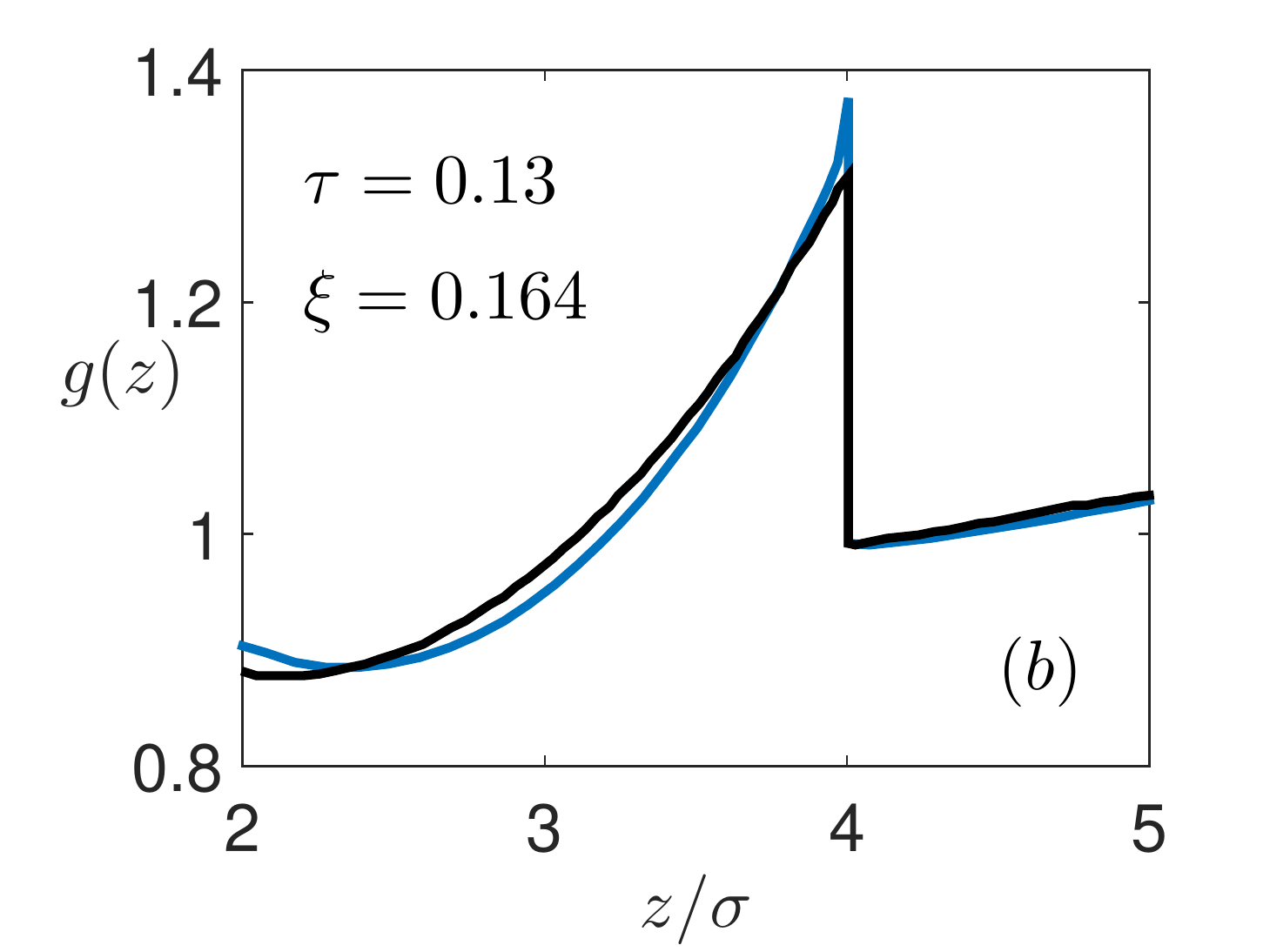}
\end{center}
\caption{(a): density profiles of SHS in confinement between two hard walls. Solid line represents our results and dots represents results of the third order perturbation theory of DFT \cite{CHGO:2002}. (b): pair correlation function of SHS for at strong stickiness $\tau=0.13$ and at packing fraction $\xi=0.164$. Black line represents our results and blue line represents results of the regulated FMT \cite{HMW:2012}.}
\label{fig:RDF}
\end{figure}

To make contact with previous results of the density functional formalism,
we switch off the LJ attraction and consider the SHS system confined by two hard walls. 
In DFT, the free-energy functional (\ref{eq:12}) is rewritten in the form,
\begin{align}
\Omega[\rho(\mathbf{r})] = F_1[\rho(\mathbf{r})] + F_0[\rho(\mathbf{r})] + \int d\mathbf{r} \big\{V_{ex}(\mathbf{r})-\mu\big\}\rho(\mathbf{r})
\end{align}
where 
\begin{align}
F_0[\rho(\mathbf{r})]=\beta^{-1}\int d\mathbf{r}\rho(\mathbf{r})\left\lbrace\ln\left[\rho(\mathbf{r})\Lambda^3\right]-1\right\rbrace
\end{align}
is the ideal part and $\Lambda$ the thermal wavelength.
The excess part $F_1$ contains the particle interactions.
To determine the density profiles and the radial distribution functions (pair correlation functions), the minimization equation is written in a form more suitable for numerical implementation,
\begin{align}\label{eq:mini_cond1}
\rho(\mathbf{r})=\rho_b\exp\left[ -\beta V_{ex}(\mathbf{r}) + c^{(1)}[\rho(\mathbf{r}),\mathbf{r}] + \beta\mu_1\right],
\end{align}
where 
\begin{align}
c^{(1)}[\rho(\mathbf{r}),\mathbf{r}]
=-\beta\frac{\delta F_1[\rho(\mathbf{r})]}{\delta\rho(\mathbf{r})},
\end{align}
is the one-point direct correlation function.
The excess chemical potential $\mu_1$ can be calculated by considering the minimization condition for the bulk fluid,
\begin{align}
\mu_1=c_0^{(1)}(\rho_b)=\mu - \beta^{-1}\ln\left(\rho_b\Lambda^3\right).
\end{align}
The radial distribution function is then evaluated using the equation for density profiles, Eq.~(\ref{eq:mini_cond1}), by fixing a particle at the origin and setting the external potential equal to the particle interaction. 
In this case, we get $g(r)=\rho(r)/\rho_b$, and write,
\begin{align}
g(\mathbf{r})=\exp\left[ -\beta \phi(\mathbf{r}) + c^{(1)}[\rho_bg(\mathbf{r})] + \beta\mu_1\right],
\end{align}
where $\phi(\mathbf{r})$ is the the particle interaction. 
For the SHS under confinement as described, we again have a quasi-1D system. 
Numerical results for the reduced density profiles $\rho(z)\sigma^3$ and pair correlation function $g(z)$ are shown Fig.~\ref{fig:RDF}.

The two confining walls are located at $z=\sigma/2$ and $z=L-\sigma/2$. 
Our results are in good agreement with previous results from a third-order perturbation theory of DFT \cite{CHGO:1997,CHGO:2002} in which the one-point direct correlation function $c^{(1)}$ has been approximated via the Percus-Yevick scheme. 
At low bulk density, the particles segregate near the bottom wall.
The density is small or vanishes at high altitude. 
At higher bulk density, the density profile is symmetric with respect to the middle of the system.
We only show the lower part. 
The pair correlation function has one SHS particle fixed at $z=0$. 

Our results are a good fit with the Rosenfeld FMT, which is known to satisfy the Percus-Yevick equation of state \cite{HMW:2012}.
However, a comparison with Monte Carlo results indicates that the Rosenfeld FMT is less accurate than White-Bear version of the FMT, which satisfies the Carnahan-Starling equation of state.
On the other hand, our approach has the advantage that it operates with arbitrary short-range interaction, whereas in the FMT, the weighted densities must be recalculated for each interaction, which is, in general, not a simple task.

There are two well known procedures to get the bulk or inhomogeneous equation of state from the free energy functional Eq.~(\ref{eq:12}). For systems with a two-body interaction, a direct and accurate method is to use the Irving-Kirkwood relation \cite{IK:1950},
\begin{align}\label{eq:ik_rel}
p(\mathbf{r})= k_BT\frac{\rho(\mathbf{r})}{m} - \frac{1}{2D}\left(\frac{\rho(\mathbf{r})}{m}\right)^2\int r'\phi'(\mathbf{r'})g(\mathbf{r'})\mathbf{r'},
\end{align}
where $D$ is the space dimensionality, $g(\mathbf{r'})\equiv g(\mathbf{r},\mathbf{r}+\mathbf{r'})$ is the pair correlation function, $m$ is the mass of the particle, and $\phi'$ is the derivative with respect to $\mathbf{r'}$ of the interaction potential. A second method is to use the expression \cite{Hill},
\begin{align}\label{eq:comp_rel}
\chi = k_BT\frac{\partial \rho(\mathbf{r})}{\partial p(\mathbf{r})}=1+\rho(\mathbf{r})\int\left[g(r')-1\right]d\mathbf{r'}
\end{align}
for the isothermal compressibility. 
If $g(\mathbf{r})$ is known exactly, then the two procedures yield exactly the same results. 
Any deviation from the exact $g(\mathbf{r})$ produces a difference between the results of the approaches \cite{TH:1963}. In DFT, the equation of state can also be calculated using the expression, 
\begin{align}
\beta p(\mathbf{r}) = \sum_{i}\rho_{i}(\mathbf{r})\frac{\delta \Phi[\rho(\mathbf{r})]}{\delta\rho_i(\mathbf{r})}-\Phi[\rho(\mathbf{r})],
\end{align}
where $\Phi[\rho(\mathbf{r})]$ is the density of free-energy functional and the sum $i$ extends over all weighted densities.

If we impose a hard wall at the origin ($z=0$), which prevents particles to be at this position, the second term in Eq.~(\ref{eq:ik_rel}) vanishes and we get a hard-wall sum rule as for the standard DFT:
\begin{align}\label{eq:cont_density}
p = k_BT\rho(0^{+}).
\end{align}
The pressure profile becomes
\begin{align}
\beta p(\mathbf{r})=-\ln\Bigl[1 - \!\int\!\! d\mathbf{r}_1\rho(\mathbf{r}_1) + \!\!\int\!\! d\mathbf{r}_1d\mathbf{r'}_1\rho^{(2)}(\mathbf{r}_1,\mathbf{r'}_1)\Bigr],
\end{align}
which for bulk fluids satisfies Eq.~(\ref{eq:cont_density}) with contact density taken from the solution of Eq.~(47).

\subsection{SHS model with sedimentation under gravity}\label{sec:sec5b} 
Here we consider a fluid bounded by two hard, horizontal walls, a distance $L$ apart, with a uniform vertical gravitational field acting on SHS colloidal particles of diameter $\sigma$ toward sedimentation.
This amounts to an external potential of the form,
\begin{align}\label{eq:ext_gravity}
V_{ex}(z)=\left\lbrace\begin{array}{ll}
+\infty & :~ z<\sigma/2,\\
mGz & :~ \sigma/2<z<L-\sigma/2,\\
+\infty & :~z>L-\sigma/2,
\end{array}\right.
\end{align}
where $mG$ is the (effective) gravitational force acting on the colloid. 
The effects of gravity become pronounced inside the colloidal regime close to the boundary with the granular regime, i.e for colloids with diameters  of several hundred nanometers.

Such systems are of great importance for studying interfacial and solvation phenomena. 
They has previously been investigated via different approaches, including the OZ integral equations formalism combined with a Percus-Yevick \cite{Rodriguez/Vicente:1996} or a hypernetted-chain\cite{Jamnik:1998} closure relation. 
They have also been studied by Monte Carlo simulations \cite{Biben/etal:1993}.

In order to facilitate contact with previous studies including Refs.~\cite{Rodriguez/Vicente:1996,Jamnik:1998}, we introduce the scaled quantities,
\begin{align}
k_1=\frac{\sigma mG}{k_BT},\qquad k_2=\frac{\sigma\mu}{k_BT}.
\end{align}
where $k_1$ controls the strength of the gravitational potential and $k_2$ controls the average colloidal density via the relation (\ref{eq:rho_normalization}). 
 
\begin{figure}[htb]
\begin{center}
\includegraphics[scale=0.27]{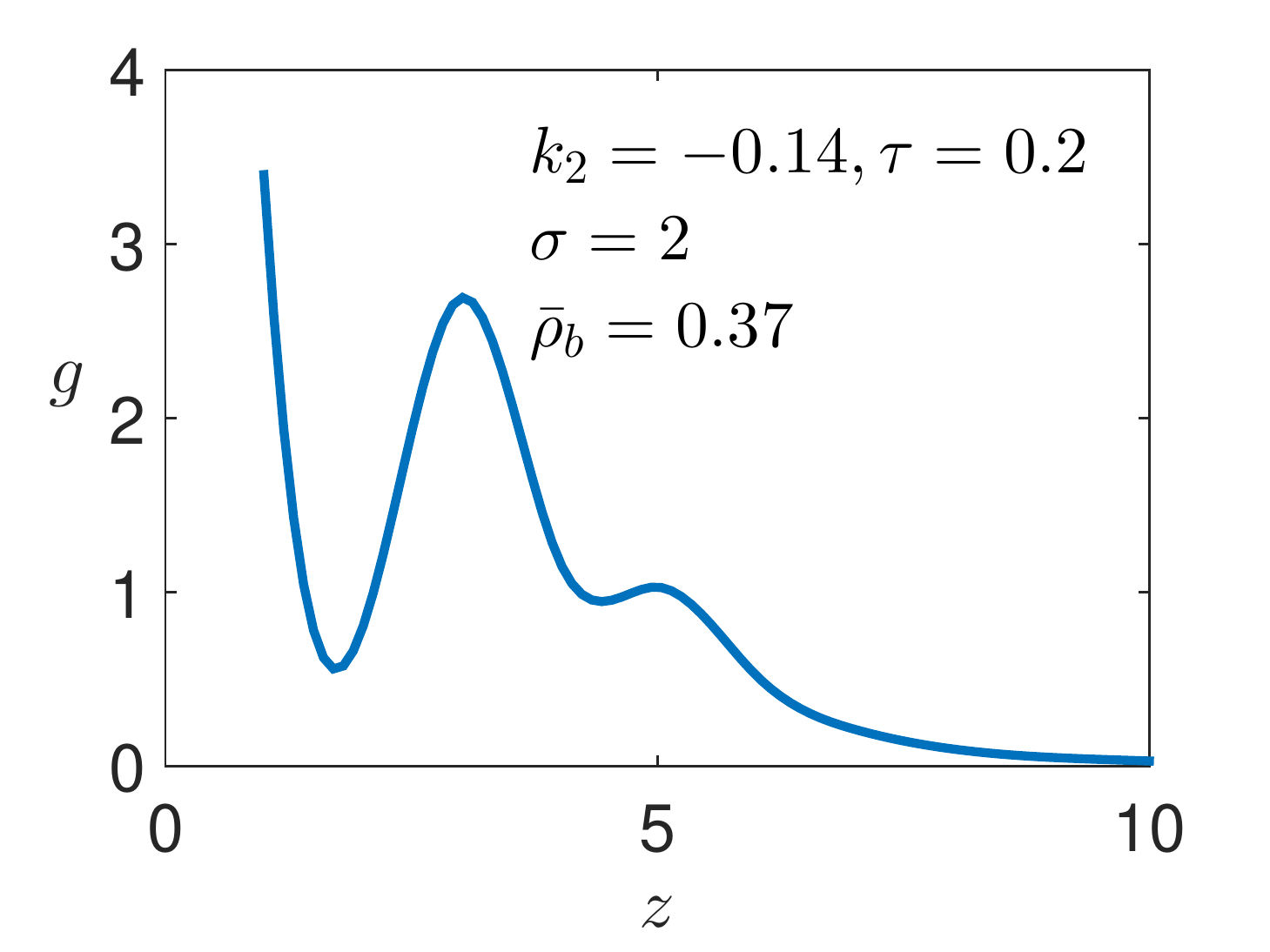}
\includegraphics[scale=0.27]{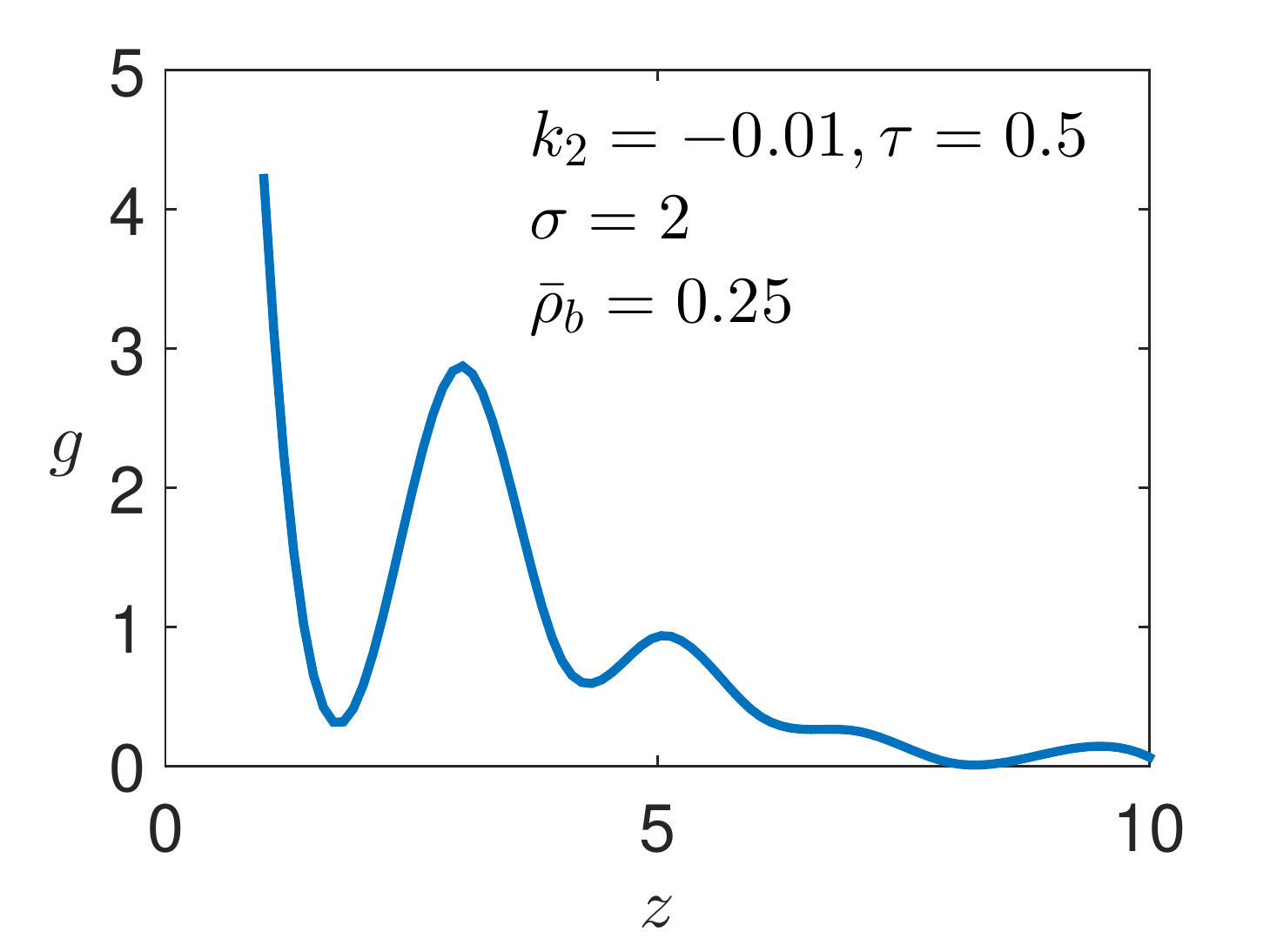}\\
\includegraphics[scale=0.27]{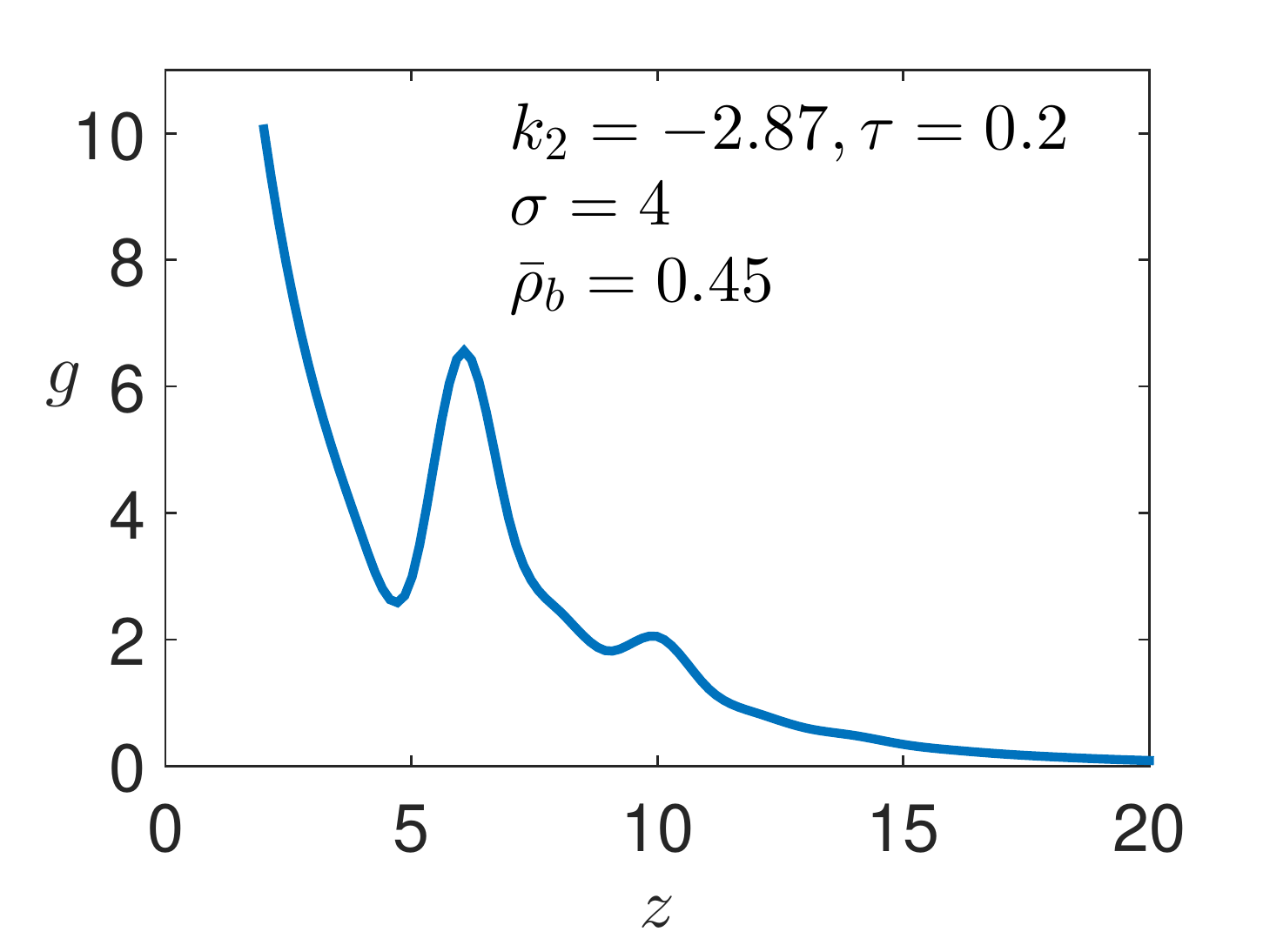}
\includegraphics[scale=0.27]{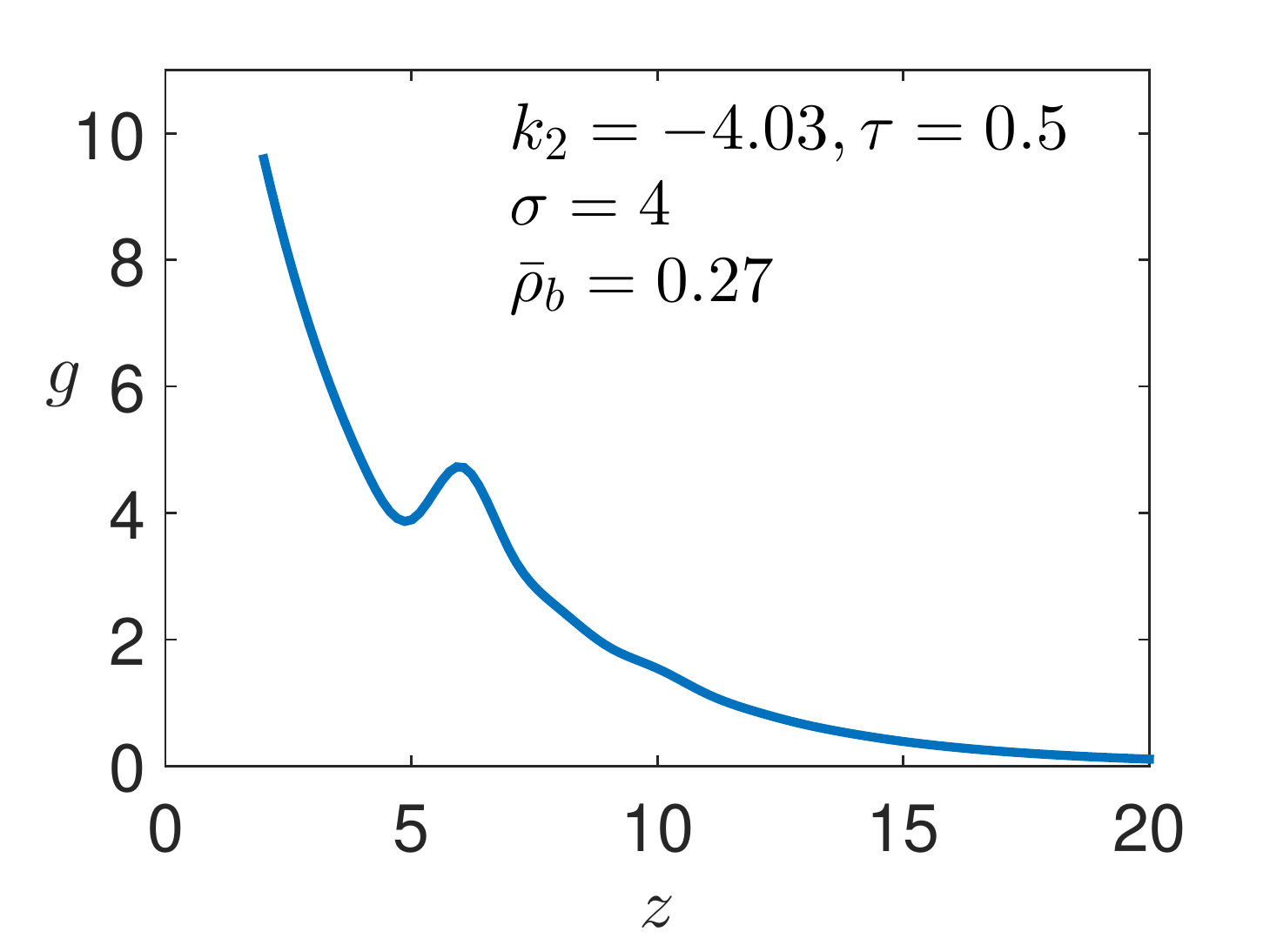}
\end{center}
\caption{One-particle distribution function $g(z)$ versus altitude $z$ for SHS with with parameter values as stated. For diameter $\sigma=2$ we have set $L=5\sigma$ and for diameter $\sigma=4$ we have set $L=11\sigma$. For all cases we have set $k_1=1.0$.}
\label{fig:g_z}
\end{figure}

\begin{figure}[t]
\begin{center}
\includegraphics[scale=0.27]{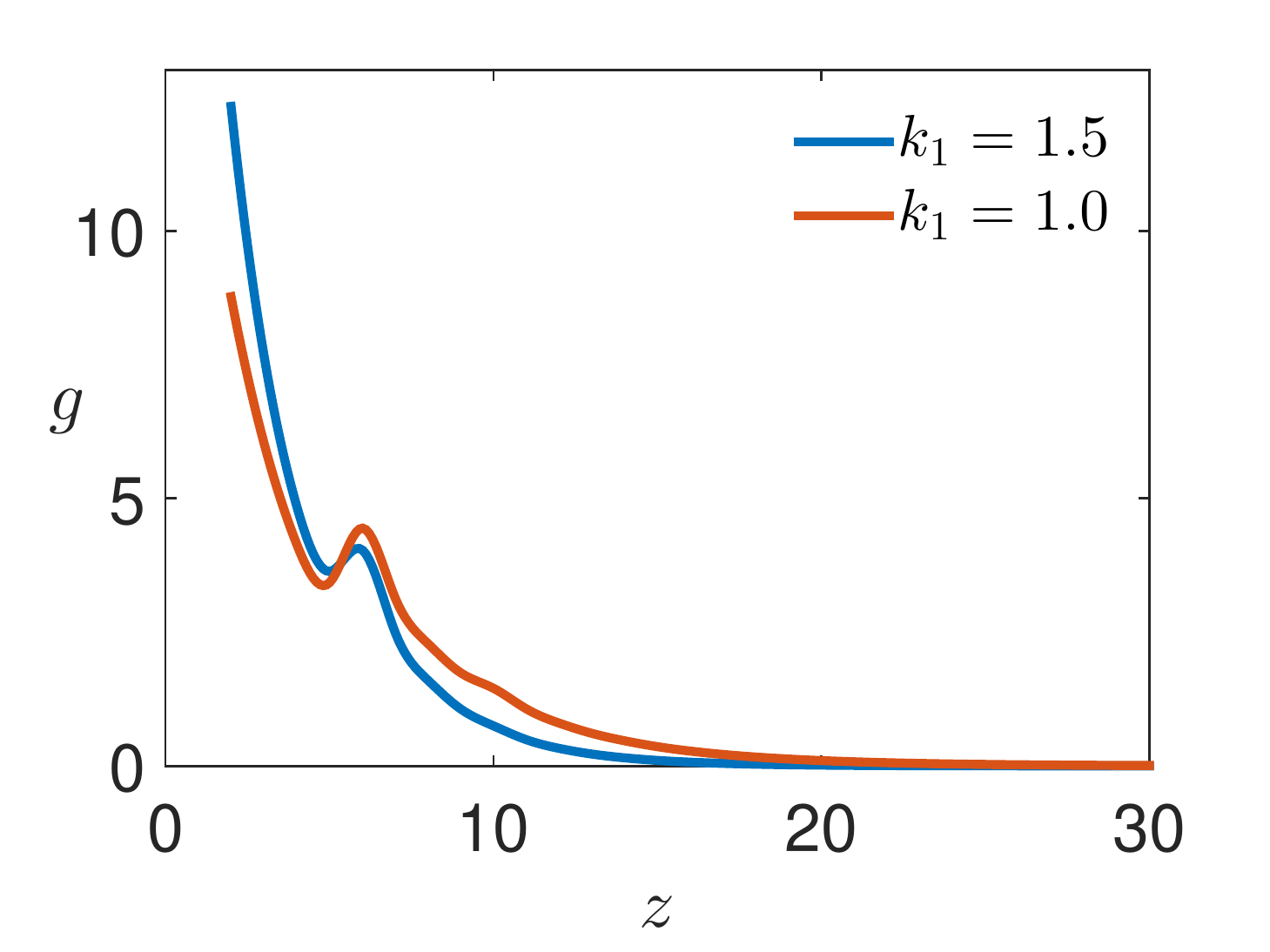}
\includegraphics[scale=0.27]{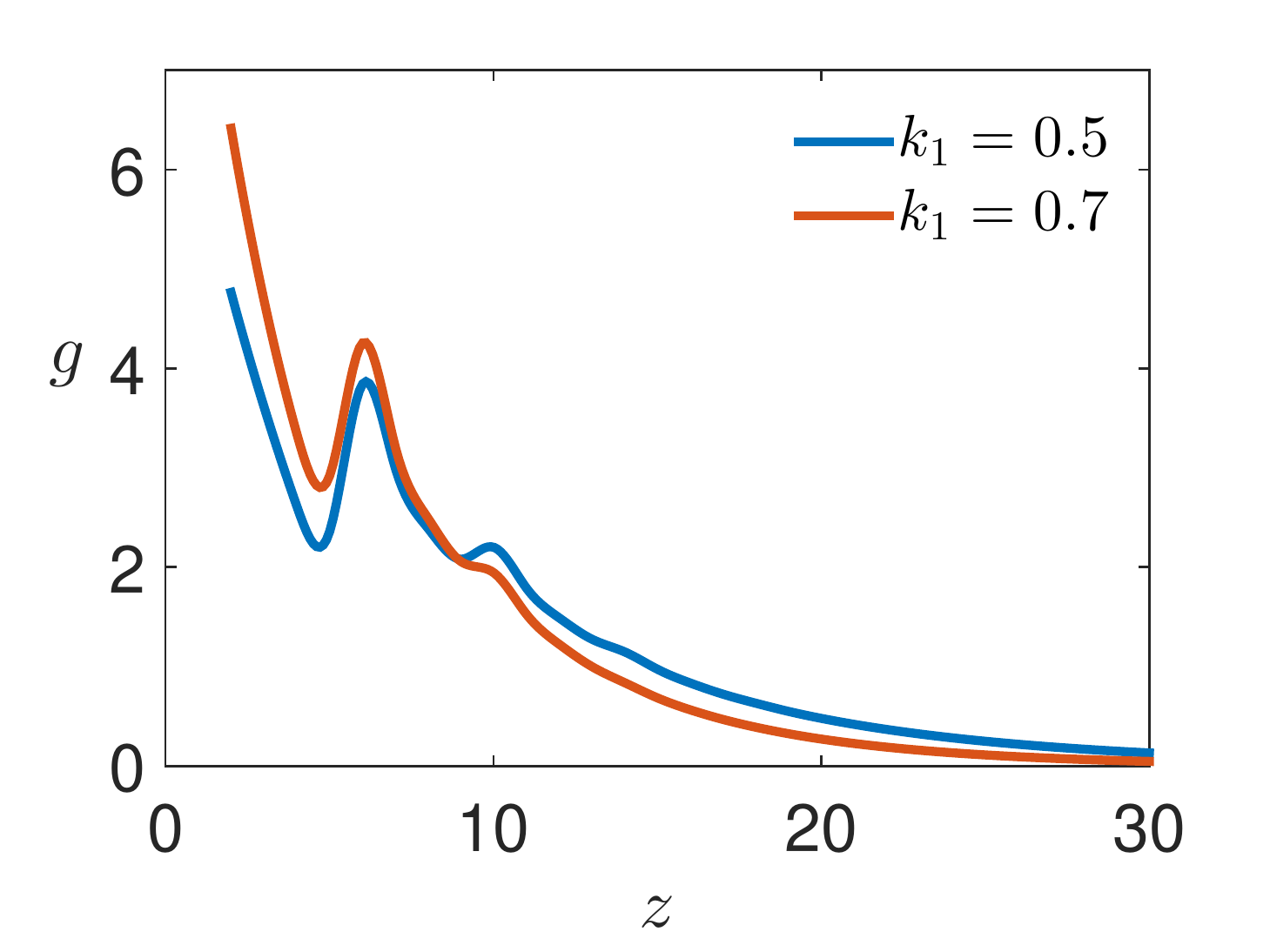}
\end{center}
\caption{One-point distribution function versus altitude for SHS with diameter $\sigma=4$ and $L=10\sigma$ at high temperature (weak gravity) and low temperature (strong gravity). The stickiness parameter is fixed to $\tau=0.7$. We have set $k_2=-3.0$.}
\label{fig:g_T}
\end{figure}

\begin{figure}[htb]
\begin{center}
\includegraphics[scale=0.5]{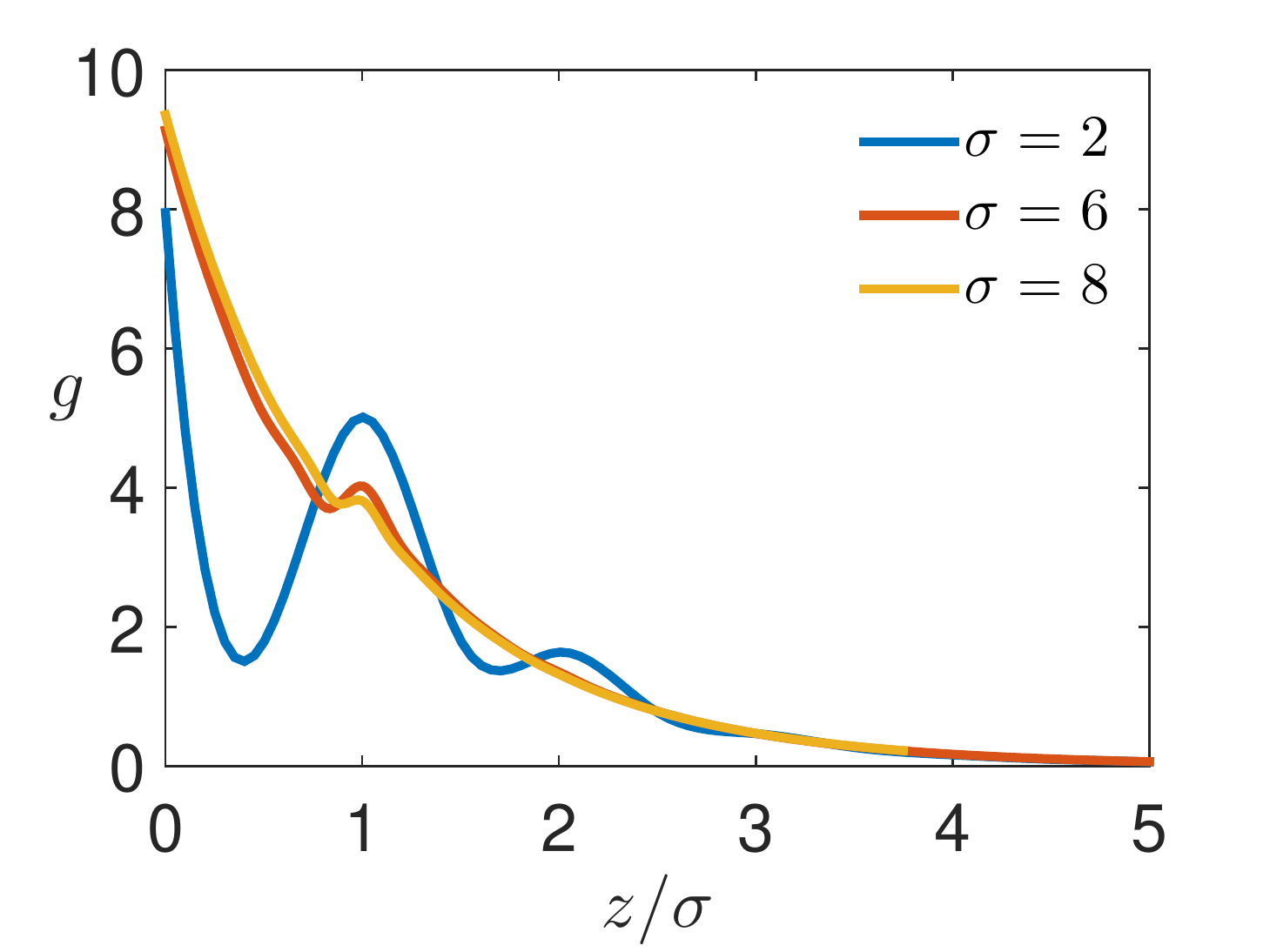}
\end{center}
\caption{One-point distribution function versus altitude for SHS with different diameters. The parameter values are set at $\bar{\rho}_b=0.111$, $L=10\sigma$, $\tau=0.7$ and $k_1=1.0$. The chemical potentials which correspond to $\bar{\rho}_b=0.111$ are $\mu=-0.1$ for $\sigma=2$, $\mu=-9.5$ for $\sigma=6$, and $\mu=-18.9$ for $\sigma=8$.}
\label{fig:g_sigma}
\end{figure}

The system of Eqs.~(\ref{eq:ddim_mu}) and (\ref{eq:32}) for the  external potential (\ref{eq:ext_gravity}) and the SHS interaction (\ref{eq:shs_boltz}) can be solved numerically to high precision using Newton, Broyden or spectral collocation methods. 
Numerical results for the one-particle distribution are shown in Figs.~\ref{fig:g_z}, \ref{fig:g_T} and \ref{fig:g_sigma}.
In our graphical representations we use the one-point distribution function,
\begin{align}
g(z)=\frac{\rho(z)}{\rho_b},\quad 
\bar{\rho}_b\doteq\rho_b\sigma^3.
\end{align}
The four panels in Fig.~\ref{fig:g_z} present variations of the one-point distribution functions for sticky hard spheres of diameters $\sigma=2$ and $\sigma=4$ under the effect gravity. 

When the gravitational energy is comparable or greater than the thermal energy, we can see pronounced oscillations in one-point distribution with amplitudes increasing when reducing the chemical potential. 
These oscillations, which arises from the hard wall combined with strong repulsive interaction of the hard spheres, are a signature of layering in the fluid that proceeds the condensation of the fluids on the surface of the wall. 
Reducing the stickiness parameter enhances the oscillations because of the dependence of the former on temperature via (\ref{eq:sw_potential}).
Parameters for the $\sigma=4$ case are taken from Ref.~\cite{Jamnik:1998} to compare with the results found there. 
Our results are manifestly in good qualitative and quantitative agreement with the those derived from the OZ approach using the Percus-Yevick approximation as a closure relation \cite{Rodriguez/Vicente:1996} and from the hypernetted chain/OZ equation \cite{Jamnik:1998}.

Increasing the effect of gravity or reducing the temperature enhances the oscillatory regime at the bottom of the wall leading to condensation of the SHS.
This is demonstrated in Fig.~\ref{fig:g_T}. 
Only at high temperature or weak gravity (meaning differential in mass density) do the colloids reach high altitude.

A noteworthy feature is the dependence of the (number) density profile on the diameter of the colloids as shown in Fig.~\ref{fig:g_sigma} at fixed bulk (number) density.
The increase in structure for decreasing diameter is quite remarkable, far from obvious.

%
\section{Conclusion}
%
We have introduced a new approach for studying the thermodynamics of interacting hard-sphere fluids in an arbitrary external field. 
The key accomplishment is a functional relation between density and PDF that works for cases of inhomogeneous fluids with general finite-range nearest-neighbor interactions. 
We have been able to construct from this functional relation explicit entropy and free-energy functionals. 

For the special 1D case of hard rods confined to a narrow channel and interacting upon contact, our results coincide with Percus' results \cite{Perc89} obtained via rigorous analysis.  
Exact solutions in explicit form can then be worked out. 
To test the strength of our approximation in higher dimensions, we have considered a system of sticky hard spheres with different external fields.

The results demonstrate that our approach is in good agreement with corresponding results inferred from commonly used approaches such as Percus-Yevick/OZ and the hypernetted chain/OZ equation and also with Monte Carlo simulation. 
Extensions of this work to mixtures of interacting fluids in arbitrary, inhomogeneous external fields are in progress. 
Extensions to molecules of different sizes and shapes, which require the inclusion of orientational degrees of freedom, are within the range of our methodology.

\acknowledgments
We thank M. Kr\"uger, A. Kl\"umper and M. Karbach for valuable
discussions.

\appendix

%
\section{Exactness of analysis in $\mathcal{D}=1$}\label{sec:appa} 
%
Here we present a proof that the functional relation (\ref{eq:6}) is rigorous for 1D systems, which implies that the expressions derived from it in Sec.~\ref{sec:sec4} are exact and ready to be evaluated for any application of choice. 
The proof uses the fact, which only holds in $\mathcal{D}=1$, that the integrals in Eq.~(\ref{eq:5A}) and (\ref{eq:5B}) are factorizable. 
The conclusion that $A=B$ follows from this attribute. 

Consider a system of $N$ hard rods (1D hard spheres) with diameter $\sigma$ confined to a region of space. 
Without loss of generality, we order the positions of rods (their centers of mass) as $x_1<x_2<\ldots < x_N$.

Introducing the functions,
\begin{align}\label{eq:app-1}
h(x)=e^{-\beta V_{ex}(x)}, \qquad f(x,y)=e^{-\beta\phi(x,y)},
\end{align}
allows us to rewrite Eqs.~(\ref{eq:5a-1})-(\ref{eq:5a-3}) in more compact form as follows:
\begin{align}\label{eq:app-2}
\rho^{(2)}(&x_i,x_{i+1})=\frac{1}{Z}h(x_i)f(x_i,x_{i+1})h(x_{i+1})
\nonumber\\
&\times\!\!\int\!\prod_{k=1}^{i-1}\!dx_kh(x_k)f(x_{k},x_{k+1})
\\
&\times \!\!\int\!\!\prod_{k=i+2}^{N}\!\!dx_k 
h(x_k)\!\!\prod_{k=i+1}^{N-1}f(x_{k},x_{k+1}),\nonumber
\end{align}
\begin{align}\label{eq:app-3}
\tilde{\rho}_0(x)&=\frac{1}{Z}\!\!\int\!\!
\prod_{k=1}^{i-1}\!\!dx_kh(x_k)\!\!
\prod_{k=1}^{i-2}\!\!f(x_{k},x_{k+1})\nonumber\\
&\times \!\!\int\!\!\prod_{k=i+2}^{N}\!\!dx_k h(x_k)\!\!
\prod_{k=i+2}^{N-1}\!\!f(x_{k},x_{k+1}),
\end{align}
\begin{align}\label{eq:app-4}
\tilde{\rho}(x,x_i)&=\frac{h(x_i)}{Z}\!\!\int\!
\prod_{k=1}^{i-1}\!\!dx_kh(x_k)f(x_{k},x_{k+1})\nonumber\\
&\times\!\!\int\!\!\prod_{k=i+2}^{N}dx_kh(x_k)\!\!
\prod_{k=i+2}^{N-1}\!\!f(x_{k},x_{k+1}),
\end{align}
\begin{align}\label{eq:app-5}
\tilde{\rho}_1(x,x_{i+1})&=\frac{h(x_{i+1})}{Z}\!\!\int\!\!
\prod_{k=1}^{i-1}\!\!dx_kh(x_k)\!\!
\prod_{k=1}^{i-2}f(x_{k},x_{k+1})\nonumber\\
&\times\!\!\int\!\!\prod_{k=i+2}^{N}dx_kh(x_k)\!\!
\prod_{k=i+1}^{N-1}\!\!f(x_{k},x_{k+1}).
\end{align}
Next we process the product in the first line of (\ref{eq:app-2}):
\begin{align}\label{eq:app-6}
\tilde{\rho}(x,x_i)&\tilde{\rho}_1(x,x_{i+1})f(x_i,x_{i+1}) \nonumber \\
&=\frac{1}{Z^{2}}h(x_i)f(x_i,x_{i+1})h(x_{i+1})\nonumber\\
&\times\!\!\int\!\prod_{k=1}^{i-1}\!\!dx_kh(x_k)f(x_{k},x_{k+1})\nonumber\\
&\times\!\!\int\!\!\prod_{k=i+2}^{N}\!\!dx_kh(x_k)\!\!\prod_{k=i+2}^{N-1}\!\!f(x_{k},x_{k+1})\\
&\times\!\!\int\!\!\prod_{k=1}^{i-1}\!\!dx'_kh(x'_k)\!\!\prod_{k=1}^{i-2}f(x'_{k},x'_{k+1})\nonumber\\
&\times\!\!\int\!\!\prod_{k=i+2}^{N}\!\!dx'_kh(x'_k)f(x_{i+1},x'_{i+2})\!\!\prod_{k=i+2}^{N-1}\!\!f(x'_{k},x'_{k+1}).\nonumber
\end{align}
When we interchange $x_k$ and $x'_k$ in (\ref{eq:app-6}), we recognize the exact relation, 
\begin{align}\label{eq:app-7}
\frac{\tilde{\rho}(x,x_i)\tilde{\rho}_1(x,x_{i+1})}{e^{\beta\phi(x_i,x_{i+1})}}=\rho^{(2)}(x_i,x_{i+1})\tilde{\rho}_0(x),
\end{align}
between the four distribution functions (\ref{eq:app-2})-(\ref{eq:app-5}).
The three ADFs (\ref{eq:rho_tilde_1})-(\ref{eq:rho_tilde_3}) simplify into
\begin{align}
&\tilde{\rho}(x,x')=\rho(x')-\left[\int_{x-\xi/2}^{x-\sigma/2}\!+\!\int_{x+\sigma/2}^{x+\xi/2}\right]dx'_1 \rho^{(2)}(x,x'_1),\label{eq:rho_tilde_1_1d}
\end{align}
\begin{align}
&\tilde{\rho}_1(x,x'')=\rho(x'') - \int_{x-\sigma/2}^{x+\sigma/2}dx'_1\rho^{(2)}(x'_1,x''),\label{eq:rho_tilde_2_1d}
\end{align}
\begin{align}
&\tilde{\rho}_0(x)\!=\!1 - \int_{x-\xi/2}^{x+\xi/2}dx'_1\rho(x'_1) 
 \label{eq:rho_tilde_3_1d}\\
& \hspace{5mm}+ \int_{x-\sigma/2}^{x+\sigma/2}dx'_1\left[
\int_{x-\xi/2}^{x-\sigma/2}\!+\!\int_{x+\sigma/2}^{x+\xi/2}\right]
dx'_2\rho^{(2)}(x'_1,x'_2).\nonumber
\end{align}
Substitution of  Eqs.~(\ref{eq:rho_tilde_1_1d})-(\ref{eq:rho_tilde_3_1d}) into Eq.~(\ref{eq:app-7}), yields the exact relation Eq.~(\ref{eq:13}) used in Sec.~\ref{sec:sec4} for the further exact analysis.

%
\section{Justification of Eq.~(\ref{eq:8}) }\label{sec:appb} 
%
Even though $\rho$ and $\rho^{(2)}$ are not independent, as is manifest in Eq.~(\ref{eq:6}), in the integral (\ref{eq:8}) with respect to $\rho^{(2)}$, we keep $\rho$ fixed. 
This course of action requires justification.
We can express the differential of the free energy functional in the form,
\begin{align}\label{eq:app_b_1}
d\Omega[\tilde{V},\phi]= \frac{\delta \Omega}{\delta\tilde{V}}\Bigg|_{\phi}\!\!d\tilde{V} + \frac{\delta \Omega}{\delta\phi}\Bigg|_{\tilde{V}}\!\!d\phi
=-\rho d\tilde{V} + \rho^{(2)}d\phi,
\end{align}
where we have used the DFT relations (\ref{eq:pdf_int_pot}). 
From the first Eq.~(\ref{eq:pdf_int_pot}), the density profiles $\rho$ can be calculated by differentiating the free energy with respect to the external potential at fixed interaction potential. 
Differentiation with respect to the interaction potential at fixed external potential produces the PDF $\rho^{(2)}$. 
Substitution of
\begin{align*}
d\left(\rho\tilde{V}\right) = \rho d\tilde{V} + \tilde{V}d\rho, \quad
d\left(\rho^{(2)}\phi\right) = \rho^{(2)}d\phi + \phi d\rho^{(2)}
\end{align*}
into Eq.~(\ref{eq:app_b_1}) yields
\begin{align}\label{eq:app_b_2}
d\Omega[\tilde{V},\phi]= -d(\rho\tilde{V}) +\tilde{V} d\rho + d(\rho^{(2)}\phi) - \phi d\rho^{(2)},
\end{align}
which we rewrite as follows:
\begin{align}\label{eq:app_b_3}
\tilde{V}d\rho - \phi d\rho^{(2)}&=d\left(\Omega[\tilde{V},\phi]+\rho\tilde{V}-\rho^{(2)}\phi\right) 
=-d(T\mathcal{S}) \nonumber \\
&=-T\frac{\delta\mathcal{S}}{\delta\rho}\Bigg|_{\rho^{(2)}}d\rho - T\frac{\delta\mathcal{S}}{\delta\rho^{(2)}}\Bigg|_{\rho}d\rho^{(2)}.
\end{align}
In consequence we can write
\begin{align}
\tilde{V} = -T\frac{\delta\mathcal{S}}{\delta\rho}\Bigg|_{\rho^{(2)}}, \quad 
\phi = T\frac{\delta\mathcal{S}}{\delta\rho^{(2)}}\Bigg|_{\rho}\label{eq:entropy_eqs_2}
\end{align}
From the second relation (\ref{eq:entropy_eqs_2}) we conclude that the entropy functional $\mathcal{S}$ follows via integration of the interaction potential $\phi$ with respect to $\rho^{(2)}$ at fixed $\rho$.




\begin{thebibliography}{100}

\bibitem{Seng13}
A. Sengupta, \emph{Topological Microfluidics} 
(Springer, New York, 2013).

\bibitem{Fanu10}
M. Fanun (Ed.), \emph{Colloids in Drug Delivery} 
(CRC Press, Boca Raton, 2010).

\bibitem{MF04a}
M. A. Miller and D. Frenkel, 
Phase diagram of the adhesive hard sphere fluid, 
J. Chem. Phys. \textbf{121}, 535 (2004).

\bibitem{MF04b}
M. A. Miller and D. Frenkel, 
Simulating colloids with Baxter’s adhesive hard sphere model
J. Phys.: Condens. Matter \textbf{16}, S4901 (2004).

\bibitem{BRP07}
S. Buzzaccaro, R. Rusconi, and R. Piazza,
“Sticky” Hard Spheres: Equation of State, Phase Diagram, and Metastable Gels,
Phys. Rev. Lett. \textbf{99}, 098301 (2007).

\bibitem{HW11}
H. Hansen-Goos and J. S. Wettlaufer,
A fundamental measure theory for the sticky hard sphere fluid,
J. Chem. Phys. \textbf{134}, 014506 (2011).

\bibitem{SZK53}
Z. W. Salsburg, R. W. Zwanzig, and J. G. Kirkwood,
Molecular Distribution Functions in a One-Dimensional Fluid,
J. Chem. Phys. \textbf{21}, 1098 (1953).

\bibitem{Tara85}
P. Tarazona,
Free-energy density functional for hard spheres,
Phys. Rev. A \textbf{31}, 2672 (1985).

\bibitem{CA85}
W. A. Curtin and N. W. Ashcroft,
Weighted-density-functional theory of inhomogeneous liquids and the freezing transition,
Phys. Rev. A \textbf{32}, 2909 (1985).

\bibitem{Evan92}
R. Evans,
Density Functionals in the Theory of Nonuniform Fluids,
in \emph{Fundamentals of Inhomogeneous Fluids}, D. Henderson (Ed.) (Marcel Dekker, New York, 1992).

\bibitem{Loew94}
H. L{\"o}wen,
Melting, freezing and colloidal suspensions,
Phys. Rep. \textbf{237}, 249 (1994).

\bibitem{CM97}
J. A. Cuesta and Y. Martinez-Raton,
Dimensional Crossover of the Fundamental-Measure Functional for Parallel Hard Cubes,
Phys. Rev. Lett. \textbf{78}, 3681 (1997).

\bibitem{TK00}
C. Tutschka and G. Kahl,
Analytic example of a free energy functional,
Phys. Rev. E \textbf{62}, 3640 (2000).

\bibitem{TCM08}
P. Tarazona, J. A. Cuesta, and Y. Martinez-Raton,
Density Functional Theories of Hard Particle Systems,
in \emph{Theory and Simulation of Hard-Sphere Fluids and Related Systems},
A. Mulero (Ed.) Lecture Notes in Physics \textbf{249} (Springer, New York, 2008).

\bibitem{HM09}
H. Hansen-Goos and K. Mecke,
Fundamental Measure Theory for Inhomogeneous Fluids of Nonspherical Hard Particles,
Phys. Rev. Lett. \textbf{102}, 018302 (2009).

\bibitem{Luts10}
J. F. Lutsko,
Recent developments in classical density functional theory,
Adv. Chem. Phys. \textbf{144}, 1 (2010).

\bibitem{selgra}
B. Bakhti, D. Boukari, M. Karbach, P. Maass, and G. M\"uller,
Density profiles of a self-gravitating lattice gas in one, two, and three dimensions,
Phys. Rev. E \textbf{97}, 042131 (2018).

\bibitem{Rose89}
Y. Rosenfeld,
Free-Energy Model for the Inhomogeneous Hard-Sphere Fluid Mixture and Density Functional Theory of Freezing,
Phys. Rev. Lett. \textbf{63}, 980 (1989).

\bibitem{RSLT96}
Y. Rosenfeld, M. Schmidt, H. L{\"o}wen, and P. Tarazona,
Dimensional crossover and the freezing transition in density functional theory,
J. Phys.: Condens. Matter \textbf{8} L577 (1996).

\bibitem{RSLT97}
Y. Rosenfeld, M. Schmidt, H. L{\"o}wen, and P. Tarazona,
Fundamental-measure free-energy density functional for hard spheres: Dimensional crossover and freezing,
Phys. Rev. E \textbf{55}, 4245 (1997).

\bibitem{Tara00}
P. Tarazona,
Density Functional for Hard Sphere Crystals: A Fundamental measure Approach,
Phys, Rev. Lett. \textbf{84}, 694 (2000).

\bibitem{LC02}
L. Lafuente and J. A. Cuesta,
Fundamental measure theory for lattice fluids with hard-core interactions,
J. Phys.: Condens. Matter \textbf{14}, 12079, (2002).

\bibitem{LC02a}
L. Lafuente and J. A. Cuesta,
Elusiveness of Fluid-Fluid Demixing in Additive Hard-Core Mixtures,
Phys. Rev. Let. \textbf{89}, 145701 (2002).

\bibitem{LC04}
L. Lafuente and J. A. Cuesta,
Density Functional Theory for General Hard-Core Lattice Gases,
Phys. Rev. Let. \textbf{93}, 130603 (2004).

\bibitem{Roth10}
R. Roth,
Fundamental measure theory for hard-sphere mixtures: a review,
J. Phys.: Condens. Matter \textbf{22}, 063102 (2010).

\bibitem{Perc89}
J. K. Percus,
Entropy of a non-uniform one-dimensional fluid,
J. Phys.: Condens. Matter \textbf{1}, 2911 (1989).

\bibitem{RV81}
A. Robledo and C. Varea,
On the Relationship Between the Density Functional Formalism and the Potential Distribution Theory for Nonuniform Fluids,
J. Stat. Phys. \textbf{26}, 513 (1981).

\bibitem{BMD00}
J. Buschle, P. Maass, and W. Dieterich,
Exact density functionals in one dimension,
J. Phys. A: Math. Gen. \textbf{33}, L41 (2000).

\bibitem{BMD00a}
J. Buschle, P. Maass, and W. Dieterich,
Wall-Induced Density Profiles and Density Correlations in Confined Takahashi Lattice Gases,
J. Stat. Phys. \textbf{99}, 273 (2000).

\bibitem{BSM12}
B. Bakhti, S. Schott, and P. Maass,
Exact density functional for hard-rod mixtures derived from Markov chain approach,
Phys. Rev. E \textbf{85}, 042107 (2012).

\bibitem{BMM13}
B. Bakhti, G. M\"uller, and P. Maass,
Interacting hard rods on a lattice: Distribution of microstates and density functionals,
J. Chem. Phys. \textbf{139}, 054113 (2013).

\bibitem{BKMM15}
B. Bakhti, M. Karbach, P. Maass, and G. M{\"u}ller,
Monodisperse hard rods in external potentials,
Phys. Rev. E \textbf{92}, 042112 (2015).

\bibitem{note1}
The sum $i<j$ is over all distinct pairs of particles and has $\frac{1}{2}N(N-1)$ terms.

\bibitem{note2}
The kinetic energy has been omitted because its contributions to all quantities under consideration are without any significant impact. 

\bibitem{PY58}
J. K. Percus and G. Yevick,
Analysis of classical statistical mechanics by means of collective coordinates,
Phys. Rev. \textbf{110}, 1 (1958).

\bibitem{LGB59}
J.M.J. van Leeuwen, J. Groeneveld, and J. de Boer,
New method for the calculation of the pair correlation function. I,
Physica \textbf{25}, 792 (1959).

\bibitem{RA79}
Y. Rosenfeld and N. W. Ashcroft,
Theory of simple classical fluids: Universality in short-range structure,
Phys. Rev. A \textbf{20}, 1208 (1979).

\bibitem{BG46}
M. Born and H. S. Green,
A General Kinetic Theory of Liquids. I. The Molecular Distribution Functions,
Proc. Roy. Soc. A \textbf{188}, 10 (1946).

\bibitem{LP66}
J. L. Lebowitz and J. K. Percus,
Mean spherical model for lattice gases with extended hard cores and continuum fluids,
Phys. Rev. \textbf{144}, 251 (1966).

\bibitem{note3}
The sum $i,j\neq i$ has $N(N+1)$ terms, where all pairs of pairs of particles are counted twice.

\bibitem{GSET96}
A. Gonis, T. C. Schulthess, J. van Ek, and P. E. A. Turchi,
A General Minimum Pronciple for Correlated Densities in Quantum Many-Particle Systems,
Phys. Rev. Lett. \textbf{77}, 2981 (1996).

\bibitem{GSTE97}
A. Gonis, T. C. Schulthess, P. E. A. Turchi, and J. van Ek,
Treatment of electron-electron correlations in electronic structure calulations,
Phys. Rev. B \textbf{56}, 9335 (1997).

\bibitem{Bakh13}
B. Bakhti,
Development of lattice density functionals and applications to structure formation in condensed matter systems,
Dissertation, Universit\"at osnabr\"uck, 2013.

\bibitem{LC05}
L. Lafuente and J. A. Cuesta,
Cluster density functional theory for lattice models based on the theory of M\"obius functions,
J. Phys. A: Math. Gen. \textbf{38}, 7461 (2005).

\bibitem{Bakhti-et-al}
B. Bakhti et al.,
Systems of interacting colloids in external field.
(unpublished).

\bibitem{BP96}
G. R. Brannock and J. K. Percus,
Wertheim cluster development of free energy functionals for general nearest-neighbor interactions in D=1,
J. Chem. Phys. \textbf{105}, 614 (1996).

\bibitem{Perc97}
J. K. Percus,
Nonuniform classical fluid mixture in one-dimensional space with next neighbor interactions,
J. Stat. Phys. \textbf{89}, 249 (1997).

\bibitem{Wert84}
M. S. Wertheim,
Fluids with highly directional attractive forces. I. Statistical thermodynamics,
J. Stat. Phys. \textbf{35}, 19 (1984).

\bibitem{Wert86}
M. S. Wertheim,
Fluids with highly directional attractive forces. III. Multiple attraction sites,
J. Stat. Phys. \textbf{42}, 459 (1986).

\bibitem{Baxter:1968}
R. J. Baxter,
Percus--Yevick Equation for Hard Spheres with Surface Adhesion.
J. Chem. Phys. \textbf{49}, 2770 (1968).

\bibitem{Hoy/OHern:2010}
R. S. Hoy and C. S. O'Hern,
Minimal Energy Packings and Collapse of Sticky Tangent Hard-Sphere Polymers.
Phys. Rev. Lett. \textbf{105}, 068001 (2010).

\bibitem{Amokrane/Regnaut:1997}
S. Amokrane and C. Regnaut,
Surface layers overlap and effective adhesion in reverse micelles: A discussion from the adhesive spheres mixture model.
J. Chem. Phys. \textbf{106}, 376 (1997).

\bibitem{Braun:2002}
F. N. Braun,
Adhesion and liquid--liquid phase separation in globular protein solutions.
J. Chem. Phys. \textbf{116}, 6826 (2002).

\bibitem{Xu/etal:2011}
Q. Xu, L. Feng, R. Sha, N. C. Seeman, and P. M. Chaikin,
Subdiffusion of a Sticky Particle on a Surface.
Phys. Rev. Lett. \textbf{106}, 228102 (2010).

\bibitem{Dreyfus/etal:2009}
R. Dreyfus, M. E. Leunissen, R. Sha, A. V. Tkachenko, N. C. Seeman, D. J. Pine, and P. M. Chaikin,
Simple Quantitative Model for the Reversible Association of DNA Coated Colloids.
Phys. Rev. Lett. \textbf{102}, 048301 (2009).

\bibitem{Stell:1995}
G. Stell,
Criticality and phase transitions in ionic fluids.
J. Stat. Phys. \textbf{78}, 197 (1995).

\bibitem{Ebner/etal:1976}
C. Ebner, W. F. Saam, and D. Stroud,
Density-functional theory of simple classical fluids. I. Surfaces.
Phys. Rev. A \textbf{14}, 2264 (1976).

\bibitem{Yatsyshin/etal:2012}
P. Yatsyshin, N. Savva, and S. Kalliadasis,
Spectral methods for the equations of classical density-functional theory: Relaxation dynamics of microscopic films.
J. Chem. Phys. \textbf{136}, 124113 (2012).

\bibitem{Marshall/etal:1996}
P. J. Marshall, M. M. Szcz{\c{e}}sniak, J. Sadlej, G. Chalasi{\`n}ski, M. A. ter Horst, and C. J. Jameson,
Ab initio study of van der Waals interaction of CO2 with Ar.
J. Chem. Phys. \textbf{104}, 6569 (1996).

\bibitem{Chizmeshya/etal:1998}
A. Chizmeshya,M. W. Cole, and E. Zaremba,
Weak Binding Potentials and Wetting Transitions.
J. Low Temp. Phys. \textbf{110}, 677 (1998).
\bibitem{Jamnik:1998}

A. Jamnik,
Suspensions of adhesive colloidal particles in sedimentation equilibrium in a planar pore.
J. Chem. Phys. \textbf{109}, 11085 (1998).

\bibitem{Rodriguez/Vicente:1996}
G. Rodríguez  and  L. Vicente,
Density profiles of colloidal suspensions in equilibrium inside slit pores.
Mol. Phys. \textbf{87}, 213 (1996).

\bibitem{CHGO:1997}
N. Choudhury and S. K. Ghosh,
Density functional theory of adhesive hard sphere fluids.
J. Chem. Phys. \textbf{106}, 1576 (1997).

\bibitem{CHGO:2002}
N. Choudhury and S. K. Ghosh,
Density Functional Theory for Baxter’s Sticky Hard Spheres in Confinement.
J. Chem. Phys. \textbf{116}, 384 (2002).

\bibitem{HMW:2012}
H. Hansen-Goos, M. A. Miller, and J. S. Wettlaufer,
Sedimentation equilibrium of a suspension of adhesive colloidal particles in a planar slit: A density functional approach.
Phys. Rev. Lett. \textbf{108}, 047801 (2012).

\bibitem{IK:1950}
J. H. Irving and J. G. Kirkwood,
The Statistical Mechanical Theory of Transport Processes. IV. The Equations of Hydrodynamics.
J. Chem. Phys. \textbf{18}, 817 (1950).

\bibitem{Hill}
T. L. Hill, \emph{Statistical Mechanics} 
(Dover Publications, 1987).

\bibitem{TH:1963}
E. Thiele,
Equation of State for Hard Spheres.
J. Chem. Phys. \textbf{39}, 474 (1963).

\bibitem{Biben/etal:1993}
T. Biben, J-P. Hansen,  and J-L. Barrat,
Density profiles of concentrated colloidal suspensions in sedimentation equilibrium.
J. Chem. Phys. \textbf{98}, 7330 (1993).



\end{thebibliography}

\end{document}